\documentclass[12pt]{article}
\usepackage{epsfig}
\begin{document}


\newcommand{\qq}{\mbox{${\rm q}\bar{\rm q}$}}
\newcommand{\qqs}{{\rm q}\bar{\rm q}}
\newcommand{\ffbar}{\mbox{${\rm f}{\bar{\rm f}}$}}
\newcommand{\ee}{\mbox{${\rm e}^+{\rm e}^-$}}
\newcommand{\eg}{\mbox{${\rm e} \gamma$}}

\newcommand{\sigee}
{$\sigma_{\mbox{e}^+ \mbox{e}^- \to \ee \mbox{Z}/\mbox{$\gamma^*$}}$}
\newcommand{\sigeg}
{$\sigma_{ \eg \to \mbox{e} \mbox{\rm{Z}} / \mbox{$\gamma^*$} }$}

\newcommand{\ww}{{\rm W}^+{\rm W}^-}
\newcommand{\zz}{\rm ZZ}
\newcommand{\mumu}{\mbox{$\mu^+\mu^-$}}
\newcommand{\tautau}{\mbox{$\tau^+\tau^-$}}
\newcommand{\lplm}{\mbox{$l^+l^-$}}
\newcommand{\eeff}{\ee\ffbar}

\newcommand{\Zg}{\mbox{${\rm Z}/\gamma^{*}$}}
\newcommand{\eeZg}{\mbox{$\ee \to {\rm \ee Z}/\gamma^{*}$}}
\newcommand{\eegee}{\mbox{$\ee \to \gamma^{*}{\rm \ee}$}}
\newcommand{\eeZ}{\mbox{$\ee \to {\rm \ee Z}$}}
\newcommand{\Zee}{\mbox{$\ee \to \ee$Z}}
\newcommand{\egez}{\mbox{${\rm e} \gamma \to {\rm eZ}$}}
\newcommand{\egezg}{\mbox{${\rm e} \gamma\to {\rm e Z}/\gamma^*$}}
\newcommand{\Zgee}{\mbox{$\ee \to {\rm (e)e}\Zg$}}
\newcommand{\WWqqen}{\mbox{$\rm{W^+W^-}\rightarrow\rm{qq}{\rm
      e}{\nu}_{\rm e}$}}
\newcommand{\qqen}{\mbox{$\qq{\rm e}{\nu}_{\rm e}$}}

\newcommand{\gstar}{\mbox{$\gamma^{*}$}}
\newcommand{\zgee}{\mbox{${\rm Z}/\gamma^{*}{\rm ee}$}}
\newcommand{\zee}{\mbox{${\rm Zee}$}}
\newcommand{\gee}{\mbox{${\rm \gamma^{*}ee}$}}

\newcommand{\MZ}{\mbox{${\rm M}_{\rm Z}$}}
\newcommand{\Zzero}{{\rm Z}$^0$}

\newcommand{\eetoee}{\mbox{$\ee\to\ee$}}
\newcommand{\eetoww}{\ee\to\ww}
\newcommand{\eetozz}{\ee\to\zz}
\newcommand{\eetoll}{\ee\to\lplm}
\newcommand{\eegammall}{\mbox{$\ee\to\gamma^*\to\lplm$}}
\newcommand{\eetogg}{\mbox{$\ee\to\ee\gamma\gamma\to\ee+{\rm hadrons}$}}
\newcommand{\eetomm}{\mbox{$\ee\to\mumu$}}
\newcommand{\eetott}{\mbox{$\ee\to\tautau$}}
\newcommand{\eetoqq}{\mbox{ $\ee\to\qq$}}

\newcommand{\gev}{\mbox{$\rm GeV$}}
\newcommand{\mev}{\mbox{$\rm MeV$}}
\newcommand{\rad}{\mbox{$\rm rad$}}
\newcommand{\pb}{\mbox{${\rm pb}$}}
\newcommand{\ipb}{\mbox{${\rm pb}^{-1}$}}
\newcommand{\Lumi}{\mbox{$\mathcal{L}$}}

\newcommand{\dedx}{{\rm d}E/{\rm d}x}
\newcommand{\necal}{N_{\rm ECAL}}

\newcommand{\qe}{\mbox{${\rm Q}_{\rm e}$}}
\newcommand{\cosmiss}{\mbox{$\cos\theta_{\rm miss}$}}
\newcommand{\coszg}{\mbox{$\cos\theta_{\qq}$}}
\newcommand{\elcosth}{\mbox{$\cos\theta_{\rm e}$}}

\newcommand{\thetacone}{\theta_{\rm cone}}
\newcommand{\econe}{E_{\rm cone}}
\newcommand{\pcone}{P_{\rm cone}}

\newcommand{\pthat}{\mbox{$\hat{p}_t$}}

\newcommand{\hats}{\mbox{$\hat{s}$}}
\newcommand{\that}{\mbox{$\hat{t}$}}
\newcommand{\hatt}{\mbox{$\hat{t}$}}
\newcommand{\hatu}{\mbox{$\hat{u}$}}
\newcommand{\pt}{\mbox{$\hat{p}_{t}$}}
\newcommand{\see}{\mbox{$\sigma_{\rm ee}$}}
\newcommand{\seg}{\mbox{$\hat{\sigma}_{{\rm e} \gamma}$}}
\newcommand{\psibar}{\mbox{$\overline{\psi}$}}
\newcommand{\me}{\mbox{$m_{\rm e}$}}
\newcommand{\mz}{\mbox{$m_{\rm Z}$}}
\newcommand{\mw}{\mbox{$m_{\rm W}$}}
\newcommand{\mqqq}{$m^2_{\rm q \bar{q}}$}
\newcommand{\mqq}{$m_{\rm q \bar{q}}$}
\newcommand{\pqq}{$p_{\rm q \bar{q}}$}
\newcommand{\Eqq}{$E_{\rm q \bar{q}}$}
\newcommand{\Dge}{\mbox{$D_{{\rm e}\gamma} (z,\hatu)_{\overline{\rm \rm EPA}}$}}

\newcommand{\Ebeam}{\mbox{$E_{\rm beam}$}}
\newcommand{\Econe}{\mbox{$E_{\rm cone}$}}
\newcommand{\Eovp}{\mbox{${E_{\rm e}}/{p_{\rm e}}$}}
\newcommand{\Elcosth}{\mbox{$\cos \theta_{\rm e}$}}
\newcommand{\Elecal}{\mbox{$E_{\rm e}$}}
\newcommand{\Evis}{\mbox{$E_{\rm vis}$}}
\newcommand{\Pvis}{\mbox{${p_{\rm vis}}$}}
\newcommand{\pmiss}{\mbox{$|p_{\rm miss}|$}}
\newcommand{\Pconv}{\mbox{${P_{\rm conv}}$}}
\newcommand{\Nct}{\mbox{${N_{\rm tracks}}$}}
\newcommand{\Nuec}{\mbox{${N_{\rm clusters}}$}}
\newcommand{\Ncone}{\mbox{${N_{\rm cone}}$}}
\newcommand{\Efwd}{\mbox{${E}_{\rm fwd}$}}
\newcommand{\Dedx}{\mbox{${\rm d}E/{\rm d}x$}} 
\newcommand{\Wdedx}{\mbox{$w({\rm d}E/{\rm d}x)$}} 
\newcommand{\gsim}{\;\raisebox{-0.9ex}
           {$\textstyle\stackrel{\textstyle >}{\sim}$}\;}
\newcommand{\lsim}{\;\raisebox{-0.9ex}{$\textstyle\stackrel{\textstyle<}
           {\sim}$}\;}
\newcommand{\rb}[1]{\raisebox{1.5ex}[-1.5ex]{#1}}
\newcommand{\raiseb}{\raisebox{1.5ex}[-1.5ex]}
\newcommand{\degree}    {^\circ}
\newcommand{\EPJ} {Eur.~Phys.~J.}
\newcommand{\PhysLett}  {Phys.~Lett.}
\newcommand{\PRL} {Phys.~Rev.~Lett.}
\newcommand{\PhysRep}   {Phys.~Rep.}
\newcommand{\PhysRev}   {Phys.~Rev.}
\newcommand{\MPLA}  {Mod.~Phys.~Lett. {\bf A}}
\newcommand{\NPB}  {Nucl.~Phys.~{\bf B}}
\newcommand{\NIM} {Nucl.~Instr.~Meth.}
\newcommand{\CPC} {Comp.~Phys.~Comm.}
\newcommand{\ZPhys}  {Z.~Phys.}
\newcommand{\etal}     {et al.}


\renewcommand{\Huge}{\huge}
\parskip12pt plus 1pt minus 1pt
\topsep0pt plus 1pt
\begin{titlepage}

\begin{center}
{\Large EUROPEAN LABORATORY FOR PARTICLE PHYSICS}
\end{center}

\begin{flushright}
CERN-EP-2001-053   \\ 10 July 2001
\end{flushright}

\bigskip\bigskip\bigskip\bigskip\bigskip
\begin{center}{\huge\bf\boldmath
 Measurement of Z/$\gamma^*$ 
production in Compton scattering of quasi-real photons
}\end{center}

\bigskip\bigskip
\begin{center}
 \large
{
The OPAL Collaboration   \\ 
}
\end{center}
\bigskip\bigskip\bigskip

\begin{abstract}
The process \eeZg\ is studied with the OPAL detector at LEP 
at a centre of mass energy of $\sqrt{s} = 189$~GeV.
The cross-section times the 
branching ratio of the \Zg\ decaying into hadrons is measured 
within Lorentz invariant kinematic limits to be 
$(1.2 \pm 0.3 \pm 0.1)$~pb for 
invariant masses of the hadronic system between
$5$~GeV and $60$~GeV and
$(0.7 \pm 0.2 \pm 0.1)$~pb for  hadronic masses above 60~GeV.
The differential cross-sections of the Mandelstam variables 
$\hat{s}$, $\hat{t}$, and  $\hat{u}$ are measured 
and compared with the predictions from the Monte Carlo 
generators grc4f and PYTHIA. 
From this, based on a factorisation ansatz, 
the total and differential cross-sections
for the subprocess \egezg\ are derived.

\end{abstract}

\bigskip\bigskip
\bigskip\bigskip

\begin{center}
{\large\bf 
(Submitted to
Eur. Phys. J. C.)}

\bigskip\bigskip

\end{center}

\end{titlepage}
\begin{center}{\Large        The OPAL Collaboration
}\end{center}\bigskip
\begin{center}{
G.\thinspace Abbiendi$^{  2}$,
C.\thinspace Ainsley$^{  5}$,
P.F.\thinspace {\AA}kesson$^{  3}$,
G.\thinspace Alexander$^{ 22}$,
J.\thinspace Allison$^{ 16}$,
G.\thinspace Anagnostou$^{  1}$,
K.J.\thinspace Anderson$^{  9}$,
S.\thinspace Arcelli$^{ 17}$,
S.\thinspace Asai$^{ 23}$,
D.\thinspace Axen$^{ 27}$
% ...
}\end{center}\bigskip
\bigskip
$^{  1}$School of Physics and Astronomy, University of Birmingham,
Birmingham B15 2TT, UK
\newline
$^{  2}$Dipartimento di Fisica dell' Universit\`a di Bologna and INFN,
I-40126 Bologna, Italy
\newline
$^{  3}$Physikalisches Institut, Universit\"at Bonn,
D-53115 Bonn, Germany
\newline
% ...
$^{  5}$Cavendish Laboratory, Cambridge CB3 0HE, UK
\newline
% ...
$^{  9}$Enrico Fermi Institute and Department of Physics,
University of Chicago, Chicago IL 60637, USA
\newline
% ...
$^{ 16}$Department of Physics, Schuster Laboratory, The University,
Manchester M13 9PL, UK
\newline
$^{ 17}$Department of Physics, University of Maryland,
College Park, MD 20742, USA
\newline
% ...
$^{ 22}$Department of Physics and Astronomy, Tel Aviv University,
Tel Aviv 69978, Israel
\newline
$^{ 23}$International Centre for Elementary Particle Physics and
Department of Physics, University of Tokyo, Tokyo 113-0033, and
Kobe University, Kobe 657-8501, Japan
\newline
% ...
$^{ 27}$University of British Columbia, Department of Physics,
Vancouver BC V6T 1Z1, Canada
\newline
% ...
Tel Aviv 69978, Israel.

\newpage
\section{Introduction}
\label{introduction}

\begin{figure}[h]
    \epsfig{figure=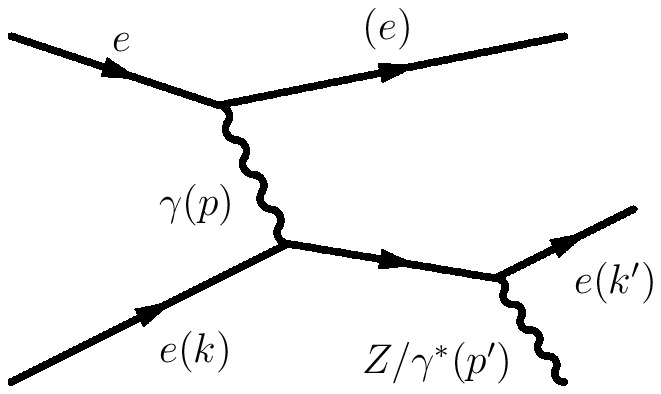,width=8.1cm}
    \epsfig{figure=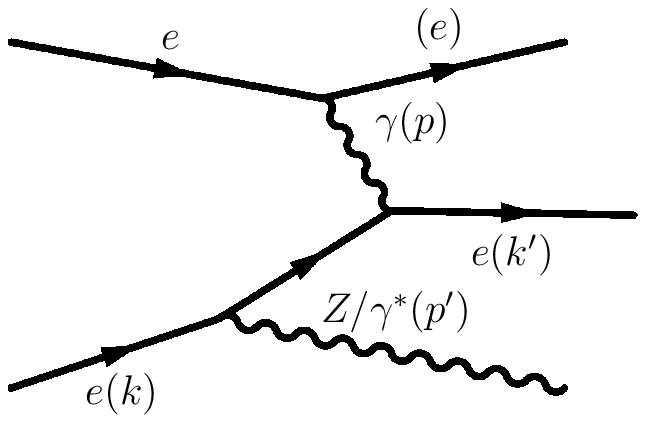,width=7cm}
    \caption{\it Diagrams for the process \Zgee.}
    \label{fig:zeefeyn} 
\end{figure}

In this paper the reaction \eeZg\ 
is studied using the OPAL detector at LEP and 
the cross-section times branching ratio for the decay of \Zg\ into hadrons, 
denoted as \see, is measured.
In this reaction a quasi-real photon is radiated from one of the beam electrons
and scatters off the other electron producing a \Zg\ as shown in
Figure~\ref{fig:zeefeyn}.
This process  was measured
for the first time~\cite{zee} with the OPAL detector.
The observable final state, (e)e\ffbar, 
consists of the scattered electron,
e, and a fermion pair, \ffbar\ , from the \Zg\ decay while the other
electron, (e), usually remains unobserved in the beam pipe due to the small 
momentum transfer squared, $|p|^2$, of the quasi-real photon. 

From the cross-section \see\ the cross-section of the subprocess
\egezg\footnote{Charge conjugation is implied throughout the paper except when
  otherwise stated.}, 
denoted by \seg, is determined.
This is the first measurement of the cross-section $\seg(\sqrt{\hats})$
for values  of $\sqrt{\hats}$ equal to or greater than the Z-mass. 
The process \egezg\
is the same as ordinary Compton scattering with
the outgoing real photon replaced by a virtual photon $\gamma^*$ or a Z.

The cross-section $\sigma_{\mathrm{ee}}$
is given by the convolution of the 
cross-section $\seg$
with the photon flux $D_{e \gamma}(z,s)$
\begin{equation}
  \sigma_{\mathrm{ee}}(s) = 
\int_0^1 {\rm d} z \; D_{{\rm e}\gamma}(z,s) \; {\rm d} \seg(\hat{s}),
  \label{eq:sepa}
\end{equation}
where $z = \hat{s}/s$.

For Z boson or $\gamma^*$ production in Compton scattering 
e$(k)\gamma(p) \to$ e$(k^\prime)$ \Zg\ $(p^\prime)$ 
of real photons ($p^2=0$), the cross-section
depends on the Mandelstam variables
$\hat{s}=(k+p)^2=(p'+k')^2$, $\hat{t}=(k'-k)^2=(p'-p)^2$ and
$\hat{u}=(p'-k)^2=(k'-p)^2$  ~\cite{gabrielli} 
\begin{equation}
 \frac{{\rm d} \hat{\sigma}_{{\rm e} \gamma}}{{\rm d} \hat{t}} \propto \frac{1}{\hat{s}^2} 
\left( \frac{\hat{u}}{\hat{s}} +
 \frac{2 m_{\rm q \bar{q}}^{2} \hat{t}}{\hat{u}\hat{s}} + 
\frac{\hat{s}}{\hat{u}} \right) .
 \label{eq:xsec}
\end{equation}
The variable 
$m_{\rm q \bar{q}}$ is the invariant mass of the quark-anti-quark pair 
the \Zg\ decays into and for $m_{\rm q \bar{q}}=0$ the well known terms 
for ordinary Compton scattering remain.

A singularity at
$\hat{u}=0$ is introduced by the virtual electron propagator in
Figure~\ref{fig:zeefeyn}(b) as 
the typical transverse momentum scale of the
scattered \Zg\ bosons is small~\cite{hagiwara}.
For incoming quasi-real photons ($p^2 \simeq 0$) in ep or \ee\ collisions, the 
dominant regulating effect for this divergence is not the electron mass,
but small, non-zero, incoming photon masses squared $p^2$. 
Via the replacement \cite{hagiwara,sjostrand}
\begin{equation}
 \hat{u} \to 
\hat{u} + p^2\frac{ m_{\rm q \bar{q}}^2}{\hat{s}} - m_{\mathrm e}^2 
 \label{eq:regulate2}
\end{equation}
in the denominator of Equation~\ref{eq:xsec}, both the photon mass
and the electron mass are included in the propagator.

A simple 
equivalent photon approximation (EPA)~\cite{salati}
\begin{equation}
D_{{\rm e}\gamma}(z,s)_{\mathrm {EPA}}
= \frac{\alpha}{2 \pi}
  \frac{1+(1-z)^2}{z} 
\bigg[ \ln \frac{s}{\me^2 } -1 \bigg] ,
  \label{eq:epa}
\end{equation}
where the integration is performed over the small photon virtualities, leads
to an effective on-shell incoming photon flux. This overestimates the 
cross-section by a factor of two~\cite{hagiwara}. 
The $p^2$ spectrum of the incoming photons either has to be retained
fully or modified to describe the process properly.
The modified EPA, denoted by $\overline{\rm{EPA}}$ is given by~\cite{hagiwara}:
\begin{equation}
  \Dge = \frac{\alpha}{2 \pi}
  \frac{1+(1-z)^2}{z} \bigg[ \ln \frac{|\hatu|(1-z)}{\me^2 z^2} -1 \bigg] 
  \label{eq:dge}
\end{equation}
In any case, the results will be sensitive to the modelling of the $p^2$
spectrum. In this paper the theoretical expectations are represented by 
Monte Carlo event generators using different approaches for obtaining the
$p^2$ spectrum of the incoming photons. These are compared with
the experimental data. 
The comparisons include the distributions
of the Mandelstam variables $\hat{s}$, $\hat{t}$, and  $\hat{u}$
as well as other characteristic variables, like
\mqq, 
and $E_e$, the energy of the scattered electron.

After giving a description of the data used for this analysis and
of the OPAL detector, a signal definition is given. Using kinematic invariants
a part of the cross-section \see\ 
is defined as signal. 
Thereafter the selection of the signal events is described and the 
total cross section \see\ within the signal definition is calculated.
Using the same selection differential cross-sections
d\see\ and d\seg\ are determined using an unfolding technique.

\section{Data and detector description}
\label{simulation}
The analysis uses 
$174.7 \pm 0.2 \; \rm{(stat.)}  \pm 0.3 $ (syst.)~pb$^{-1}$ of data
collected during $1998$ with the 
OPAL detector at LEP at a
centre of mass energy of $\sqrt{s} \simeq 189$~GeV.
A detailed description of the OPAL detector may be found 
elsewhere~\cite{detector} and only a short description is given here.
The central detector consists of
a system of tracking chambers
providing charged particle tracking
over 96\% of the full solid 
angle\footnote
   {The OPAL coordinate system is defined so that the $z$ axis is in the
    direction of the electron beam, the $x$ axis 
    points towards the centre of the LEP ring, and  
    $\theta$ and $\phi$
    are the polar and azimuthal angles, defined relative to the
    $+z$- and $+x$-axes, respectively. The radial coordinate is denoted
    as $r$.}
inside a 0.435~T uniform magnetic field parallel to the beam axis. 
It is composed of a two-layer
silicon microstrip vertex detector, a high precision drift chamber,
a large volume jet chamber and a set of $z$ chambers measuring 
the track coordinates along the beam direction. 
A lead-glass electromagnetic (EM)
calorimeter located outside the magnet coil
covers the full azimuthal range with excellent hermeticity
in the polar angle range of $|\cos \theta |<0.82$ for the barrel
region  and $0.81<|\cos \theta |<0.984$ for the endcap region.
The magnet return yoke is instrumented for hadron calorimetry 
and consists of barrel and endcap sections along with pole tip detectors that
together cover the region $|\cos \theta |<0.99$.
Four layers of muon chambers 
cover the outside of the hadron calorimeter. 
Electromagnetic calorimeters close to the beam axis 
complete the geometrical acceptance down to 24 mrad, except
for the regions where a tungsten shield is designed to protect
the detectors from synchrotron radiation.
These calorimeters include 
the forward detectors (FD) which are
lead-scintillator sandwich calorimeters and, at smaller angles,
silicon tungsten calorimeters~\cite{ref:SW}
located on both sides of the interaction point.
The gap between the endcap EM calorimeter and the FD
is instrumented with an additional lead-scintillator 
electromagnetic calorimeter,
called the gamma-catcher.
The tile endcap~\cite{OPAL-TE} scintillator arrays are 
located at $0.81 < |\cos \theta | < 0.955$ behind the pressure bell and 
in front of the endcap ECAL. 
Four layers of scintillating tiles~\cite{OPAL-TE} 
are installed at each end of the detector and 
cover the range of $0.976 < |\cos \theta | < 0.999$.

\section{Signal definition and event simulation}
\label{cha:sigdef}

\subsection{Signal definition}

The predominant signature of the signal process in the final state
(e)eqq is one electron, 
two hadronic jets from the \Zg\ decay and large missing 
momentum in the direction of the beam pipe due to the escaping electron.
The cross-section is peaked at low $|\hat{u}|$ where the scattering angle of 
the signal electron is large, i.e.\ 
in the 
backward\footnote{The forward direction is defined by the initial 
                  direction of the 
                  electron radiating the \Zg .}
direction, 
and its energy is small.
Furthermore, $|\hat{u}| \rightarrow 0$ implies that the \Zg\ is emitted close
to the forward direction. As a consequence a huge part of the cross-section 
lies outside of the acceptance of the OPAL detector;
therefore the signal is defined  
within kinematic limits.

The process \eeZg\ and its subprocess \egezg\
can also be measured at a future 
\ee\ linear collider~\cite{hagiwara,choi}. 
There this process will be the dominant source of real Z production.
Furthermore this subprocess can be observed in ep 
collisions~\cite{gabrielli,cornet} where the
beam proton emits a bremsstrahlung photon.
The relevant quantity for the e$\gamma$ collision is $\sqrt{\hat{s}}$, 
the centre of mass energy in the e$\gamma$ rest-frame. 
In order to be able to compare the results of this analysis with the
measurements of other experiments Lorentz 
invariant quantities are used for the definition of the signal.
This is in contrast to~\cite{zee} where 
the signal was defined by the geometrical acceptance of 
the detector rather than by Lorentz invariant quantities. Consequently 
the results from the previous paper cannot be compared directly with the ones
presented here.

\begin{figure}[h]
    \epsfig{figure=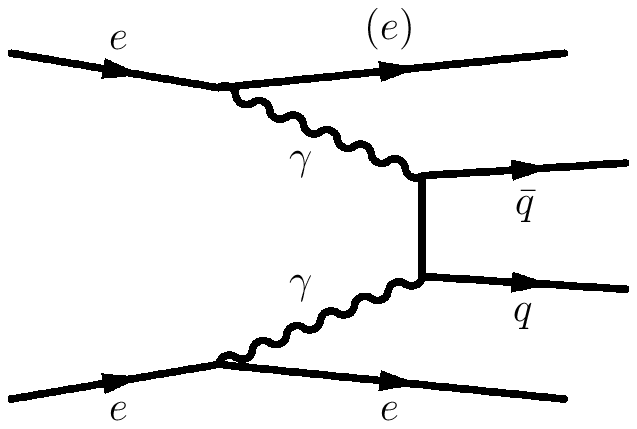,width=8.1cm}
    \vspace{1cm}
    \epsfig{figure=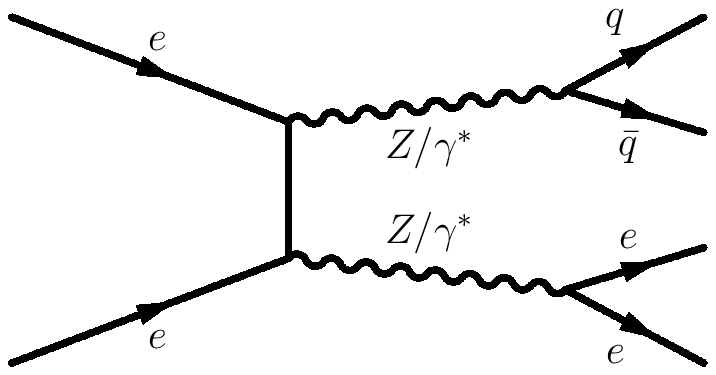,width=6.7cm}
    \caption{\it Further diagrams leading to the final state eeqq.
    On the left is the multiperipheral diagram and 
    on the right is the conversion diagram.}
    \label{fig:othfeyn} 
\end{figure}

The signal is defined by the two diagrams shown in Figure~\ref{fig:zeefeyn}
within additional kinematic limits as detailed below.
Further processes like the ones shown in~Figure~\ref{fig:othfeyn}
leading to an (e)eqq final state
are treated as background even if they satisfy our signal definition.

The Feynman diagrams for $t$-channel Bhabha scattering with initial or final
state radiation are identical to the \eegee\ Compton scattering diagrams. 
While the Bhabha 
events with initial-state radiation of a virtual photon
correspond to the $u$-channel \gee\ diagram,  Bhabha events  with final state
radiation are equivalent to the $s$-channel \gee\ events.
The cross-section for Bhabha scattering diverges for
$p^2 \rightarrow 0$. This divergence is regulated by a finite $p'^2$
of the radiated $\gamma^*$ but it still causes the cross-section to be
largely peaked at small $|\hat{t}|=|(p'-p)^2|$.
Bhabha scattering with $\gamma^*$ radiation may best be characterised
by two energetic electrons (small $p^2$ of the exchanged photon)
and a preferably low-momentum $\gamma^*$ (small $p'^2$)
in the \ee\  centre of mass system, leading to small $|\hat{t}|$.
In the observable phase space  of the
e$^+$e$^- \to$ e$^+$e$^-$Z$/\gamma^*$ process on the other hand,
the energies of the incoming photon ($p^2$) and outgoing \Zg\ ($p'^2$)
are sizeable and their momenta prefer opposite directions,
leading to large negative values of $\hat{t}$.
We therefore require the absolute value of the kinematic invariant $|\hat{t}|$
to be larger than $500$~GeV$^2$ to define our signal.

The square of the four-momentum transfer of the quasi-real photon,
$|p^2|$, is required to be less than
$10$~GeV$^2$ to ensure that the electron emitting the quasi-real photon
stays within the beam-pipe. 
As $|p^2|$ is usually small, this requirement does not reduce
the cross-section by much. 
In order for the $\overline{\rm{EPA}}$ from Equation~\ref{eq:dge}
to provide correct results the virtuality of the quasi-real 
photon needs to be the smallest virtuality in the process. 
This is guaranteed by requiring  $| \hatu |$ 
of the  electron in Figure~\ref{fig:zeefeyn}b)
to be larger than  $10$~GeV$^2$, the cut value on $|p^2|$.
This cut mainly rejects events which would be 
very difficult to select because either the energy of the scattered electron
is small or the scattering angle is very close to the beam direction.

In order to avoid the
region of hadronic resonances with all its uncertainties in the simulation of
the spectrum we require the square of the invariant mass of the \Zg, \mqqq, 
to be greater than $25$~GeV$^2$.
The kinematic limits for the signal are summarised in table~\ref{tab:sig}.

\begin{table}[htb]
\begin{center}
\begin{tabular}{|l|c|} 
\hline
 Angle and energy of the signal electron: 
 & $|\hat{t}| \ge 500 ~\gev^2 $ \\
\hline
 Mass square of the quasi-real photon:
 & $|p^2| \le 10 ~\gev^2 $ \\
\hline
Virtuality of electron:
& $|\hatu|>10~\gev^2$  \\
\hline
 Mass square of the \qq ~system:
 &  \mqqq $ \ge 25 ~\gev^2 $ \\
\hline
\end{tabular}
\caption{\it Cuts used for the definition of the signal}
\label{tab:sig}
\end{center}
\end{table}

\begin{figure}[h]
  \centerline{
    \epsfig{figure=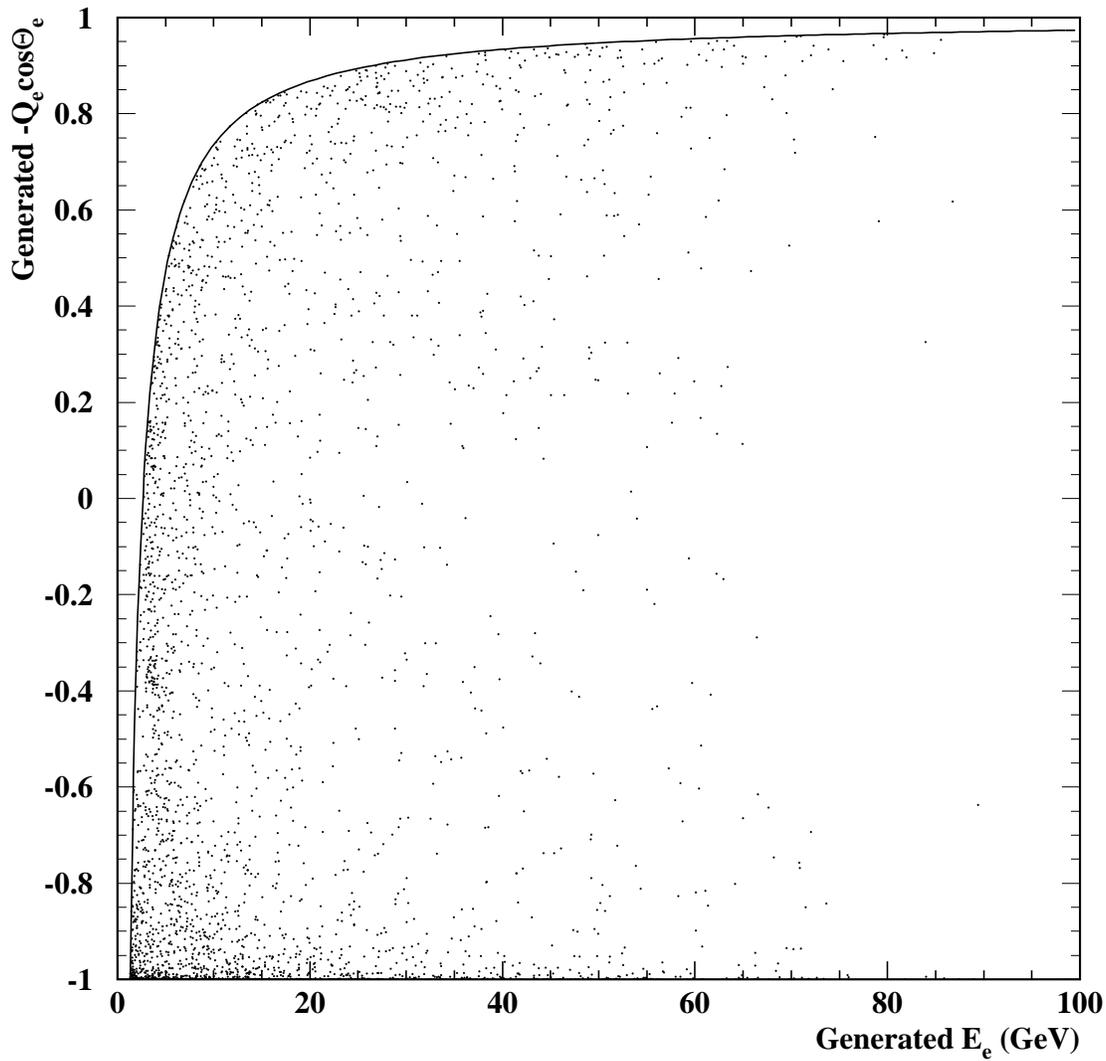,width=1.\textwidth}}
    \caption{\it Distribution of the charge weighted cosine of the 
      scattering angle of the
      electron $-$\qe\elcosth\ versus its energy \Elecal\ as predicted
      by the grc4f signal Monte Carlo at generator level.
      The solid line corresponds to
      $|\hat{t}| = 500 \; \gev^2$.} 
    \label{fig:correl} 
\end{figure}

The interdependence of the Mandelstam variables is given by
\begin{eqnarray}
\hat{t}& = &-\frac{1}{2} (\hat{s}- m_{\rm q \bar{q}}^2) \; (1-\cos\theta^*)  \\
\hat{u}& = &-\frac{1}{2} (\hat{s}- m_{\rm q \bar{q}}^2) \; (1+\cos\theta^*)
\label{eq:ut}
\end{eqnarray}
where $\cos\theta^*$ is the scattering angle of the \Zg\ 
with respect to the e$\gamma$ axis 
in the e$\gamma$ rest-frame.
Defining the kinematic region of the signal within $|\hat{t}| \ge 500 ~\gev^2$
and   \mqqq $ \ge 25 ~\gev^2 $ is an effective cut on the centre of mass
energy in the e$\gamma$ rest-frame at $\sqrt{\hat{s}} \ge 22.9$~GeV.
The kinematic invariant $\hat{t}$ is
completely determined by the four-momentum of the electron and $\hat{u}$ is
determined by the hadronic decay products of the \Zg. 
Neglecting the mass of the electron one obtains
\begin{eqnarray}
\hat{t} & = &- 2 \; E \Elecal \; (1+\qe\elcosth)  \\
\label{eq:te}
\hat{u} & = &- 2 \; E E_{\qqs} \; 
\left( 1 - \frac{p_{\qqs}}{E_{\qqs}}  \qe \cos\theta_{\qqs} \right)+ 
m_{\rm q \bar{q}}^2 ,
\label{eq:utlab}
\end{eqnarray}
with $E$ being the energy of the beam electrons, 
$\Elecal $, $\elcosth $ and $\qe$
the energy, scattering angle and the sign of the charge of the electron,
$E_{\qqs}$ and $p_{\qqs}$ the energy and momentum of the hadronic system.
The cut on $\hat{t}$ is therefore a cut in the $[\Elecal, -\qe \elcosth]$ plane
as depicted in Figure~\ref{fig:correl}. Since $\elcosth \ge -1$ there is a
hard cutoff on $\Elecal \ge 1.3$~GeV.

\subsection{Signal Simulation}

For the generation of the \eeZg\ signal events 
two different Monte Carlo generators, grc4f~\cite{grace} 
and PYTHIA 6.133~\cite{pythia}, are used.

The grc4f Monte Carlo
generator is linked to GRACE, an automatic Feynman diagram
computation program.
The total and differential cross-sections are obtained from a phase space
integration of the matrix element, which is calculated from all diagrams
corresponding to a given initial and final state. 
All fermion masses are non zero and
helicity information is propagated down to the final state particles.
Also  a subset of diagrams can be chosen. Here only diagrams according
to the signal definition have been used.
A sample of events corresponding to about 30 times the data luminosity 
has been analysed.

In PYTHIA the cross-section is calculated according to 
Equation~\ref{eq:xsec} including the regularisation given in 
Equation~\ref{eq:regulate2} to avoid the divergency in the matrix element.
In contrast to grc4f the matrix element for the process \egezg\ 
and the modified EPA as given in Equation~\ref{eq:dge}
are being used.
In PYTHIA a cutoff on ${\hat p}_{t}$, the transverse momentum of the \Zg\ 
with respect to the axis of motion of the electron and the photon
in the e$\gamma$
rest-frame, is applied.
The default cutoff of $1$~GeV has been removed in order to
include the full phase space.
This has been made possible by introducing
the new regularisation of Equation~\ref{eq:regulate2} into PYTHIA 6.133.
A further replacement is made to ensure  the cross-section does not 
become negative:
\begin{equation}
 \hat{t} \to \hat{t} - \frac{p^2}{m_{\rm{q \bar{q}}}^2}\hat{t} 
 \label{eq:regulate3}
\end{equation}
A sample of events corresponding to approximately 11 times the data 
luminosity has been used.

For both Monte Carlo generators 
parton showers and hadronization of the final quarks are performed by 
JETSET~\cite{pythia}
with parameters tuned to the OPAL data \cite{bib-OPALtune}.

Within the kinematic limits defined above
the cross-section \see\
is predicted by grc4f to be
$(1.77 \pm 0.02)$~pb, 
while the corresponding value
from PYTHIA is $(1.92 \pm 0.03)$~pb. The errors are statistical only.

\begin{figure}[h]
  \centerline{
    \epsfig{figure=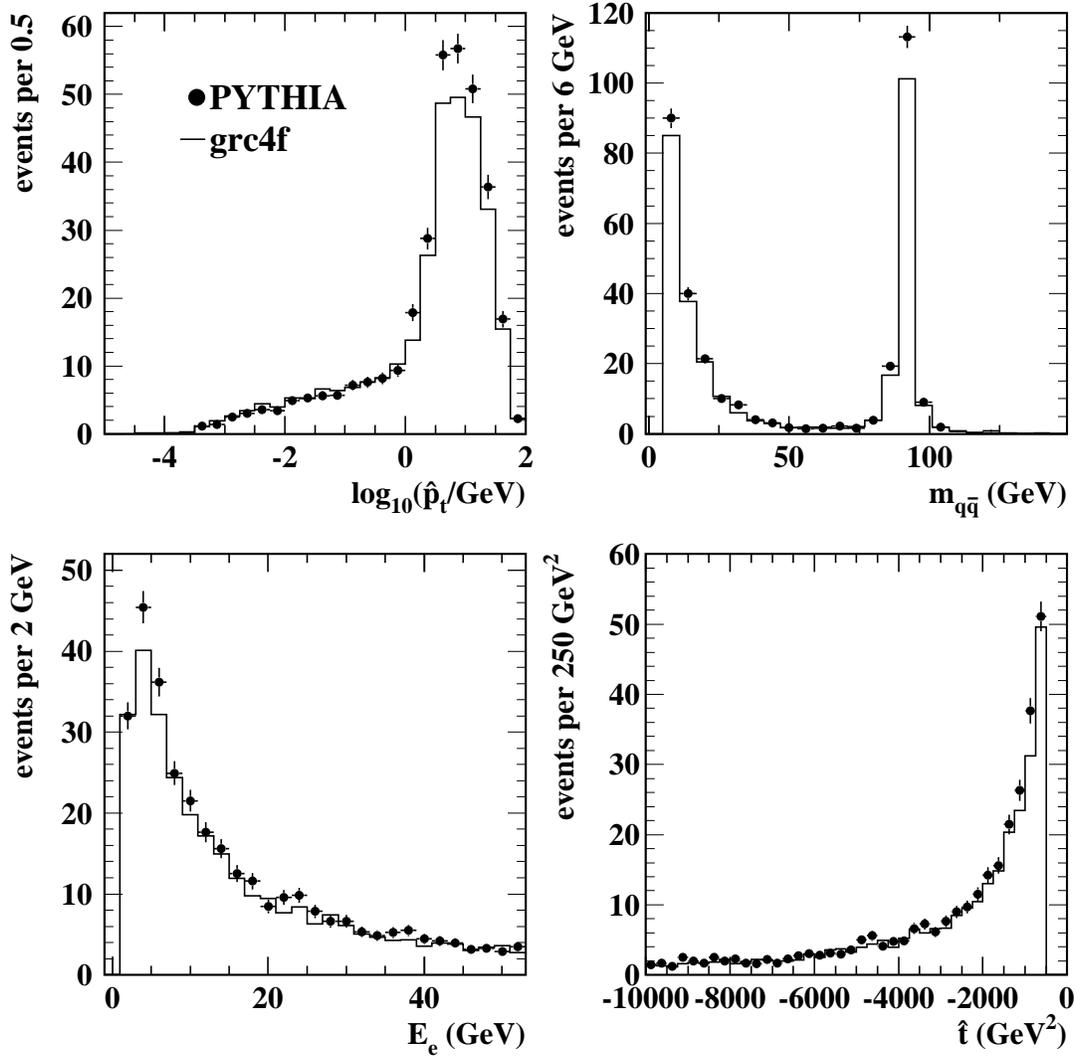,width=16cm}}
    \caption{\it Distribution of the transverse momentum \pthat\ 
        of the $\Zg\ $in the e$\gamma$
        rest-frame, the hadronic mass, \mqq, the electron energy, E$_e$,
        and \that\ at generator level for the signal for all events 
        fulfilling the signal definition. 
        The histograms show the distributions for the grc4f sample and
        the points for the PYTHIA sample.
        Only the statistical errors of the PYTHIA sample are shown.
        The distributions are normalised to the data luminosity.
        }
    \label{fig:pthat} 
\end{figure}

Figure~\ref{fig:pthat} shows the distribution of \pthat, \mqq, $E_{\mathrm e}$ 
and 
\that\ on generator level for the two Monte Carlo samples after applying the
signal definition. 
PYTHIA predicts a higher cross-section, mainly at small electron energies and
large values of 
\pthat.
In the \mqq\ distribution the contributions from the \gstar\
and the Z are well separated.

\subsection{Background Simulation}

The main contribution to the background comes from two-photon hadronic
processes, 
$\ee \to  \ee +$ hadrons. 
These events are divided into three subsets, depending on the 
virtualities\footnote{The momentum transfer squared, 
  $q^2 \equiv -Q^2$, in 
  two-photon processes is by definition identical to $\hat{t}$ 
  in our
  signal process and $p^2 \equiv -P^2$ is identical to our $p^2$.},
$q^2$ and $p^2$,
of the photons and consequently the number of beam electrons
observed (``tagged'') in the detector.
For the
low momentum transfer
processes (``untagged''), both $q^2$ and $p^2$ are small;
these are simulated using PYTHIA~5.7.
Where the momentum transfer of one of the
photons is large (``single tagged''), i.e. $q^2$ is large and $p^2$ small, 
HERWIG~\cite{herwig} is used. 
The PHOJET~\cite{phojet} generator is used for
the processes where both $q^2$ and $p^2$ are large 
(``double tagged''), i.e. $4.5$~GeV$^2 < p^2, q^2 < 50$~GeV$^2$,
which only gives a very small contribution to the background.
Two-photon production of \ee\lplm\ is simulated by the
VERMASEREN~\cite{vermaseren} Monte Carlo generator.
The different Monte Carlo samples are added to provide a complete two-photon
sample without double counting. 

Four-fermion processes like conversion and bremsstrahlung diagrams, 
except for the multiperipheral (two-photon)
processes, are studied using grc4f. As the studied process
\eeZg\ is also contained in this class a signal definition is applied
(Section~\ref{cha:sigdef}) to classify events either as signal or background.

Multi-hadronic background \eetoqq\ is simulated using PYTHIA.
As a cross-check sample YFS3FF~\cite{YFS} has been used.
Other background processes involving two fermions in the final state are
evaluated using KORALZ~\cite{koralz} for \eetomm\ and \eetott\ and
BHWIDE~\cite{bhwide} and TEEGG~\cite{teegg} for $\eetoee (\gamma)$.

The integrated luminosity of each of these samples corresponds to
at least 5 times the data luminosity, except for the two-photon samples,
which correspond to at least the same as the data 
luminosity.
All Monte Carlo samples are passed through the OPAL detector simulation
~\cite{gopal} 
and were subjected to the same reconstruction code as the data.

The contribution from processes leading to an (e)eqq final state
fulfilling the kinematic cuts of the signal definition 
but stemming from other diagrams than the ones shown in 
Figure~\ref{fig:zeefeyn} has been calculated.
This (e)eqq background includes
processes from multiperipheral and conversion diagrams 
shown in Figure~\ref{fig:othfeyn}.
For tagged two-photon events, the cross-section 
within our kinematic limits predicted by the HERWIG Monte Carlo simulation is
$\sigma = (0.88 \pm 0.03)$~pb. For the conversion processes grc4f predicts a
cross-section of $\sigma = (0.23 \pm 0.02)$~pb within the defined kinematic 
region.

\section{Event preselection}
\label{preselection}
The preselection is designed to extract events with
two jets and one isolated electron.
Only tracks and clusters which satisfy standard quality 
criteria are considered. 
  An algorithm~\cite{MT} 
  which corrects for double counting of energy between
  tracks and calorimeter clusters has been used to determine the missing
  energy and momentum.

\begin{itemize}
\item From the hadronic \Zg\ decay two jets are  expected
  in the signal events. For that reason the sum of tracks  and
  electromagnetic calorimeter clusters  unassociated
  to tracks is required to be greater 
  than $5$. 
\item
  All the tracks in the event with an associated 
  electromagnetic cluster of energy more than $1$~GeV 
  are considered as electron candidates.
  The ratio of cluster energy to track momentum 
  is required to fulfil $\Eovp \ge 0.7$.
  The specific energy loss \Dedx\ of the track in the jet chamber must 
  be consistent with the one for electrons.
  Rejection of electrons originating from photon conversions is implemented
  using the output of a dedicated artificial neural network~\cite{idncon}. 
  As an isolation criterion no additional track 
  in a cone of $0.25$~rad half opening angle 
  around the candidate electron track is allowed.
  After subtracting the energy of the candidate
  the energy deposit in this cone must be less than $10$~GeV.
  If more than one electron candidate
  track satisfies these criteria, the one
  with the smallest additional energy deposit within the cone
  is  selected.

  The charge
  of the candidate has to be consistent with the direction of the missing
  momentum 
  i.e. $\qe \cosmiss \ge 0$, 
  where \qe\ is the charge of the track considered as an electron candidate
  and $\theta_{\mathrm{ miss}}$ the polar angle
  of the missing momentum.

\item Following
  the signal definition it is required 
  that the invariant mass squared
  of the hadronic system is larger than $25$~GeV$^2$. The energies and momenta
  of the two jets are obtained from a kinematic fit.
  The kinematics of the event has to be consistent with
  a topology of two jets and two electrons,
  with one of the electrons going unobserved along the beam pipe. 
  The reconstruction of the jets is performed by
  the $k_{\perp}$ ``Durham'' \cite{durham} jet-finding algorithm.  
  The four-vector of the unobserved electron is assumed to be 
  $(0,0,p_{z\,{\rm unobs}},E_{\rm unobs})$, with 
  $|p_{\rm unobs}|=E_{\rm unobs}$.
  Energy and momentum conservation are used in the fit within the experimental 
  errors of the
  two jets and the signal electron candidate. An error of $10$ mrad has been
  assigned to the direction of the momentum of the untagged electron.
  The probability of the kinematic fit has to be larger than $10^{-6}$.

\item 
  The e$\gamma$ centre of mass energy $\sqrt{\hat{s}}$
  is obtained from the fitted energies and
  momenta of the two jets and the isolated electron.
  In the signal definition there is an effective 
  cutoff on $\sqrt{\hat{s}}$ at $22.9$~GeV. 
  Taking into account the resolution in $\sqrt{\hat{s}}$,
  a cut $\sqrt{\hat{s}} \ge 25$~GeV is applied.
\item 
  Following the signal definition 
  $|\hat{t}|$ has to be greater than  $500 \;\gev^2$.
  A cut $|\hat{t}| > 500 \;\gev^2$,
  where $\hat{t}$ is calculated according to equation~\ref{eq:te},
  is applied.
\item 
  In order to reject Bhabha events the contribution of
  the three highest energetic electromagnetic clusters to the total observed 
  electromagnetic
  energy is required to be less than $93 \%$.
  Furthermore at least one track with a  
  specific energy loss \Dedx\ in the central tracking chamber~\cite{dedx}
  not being consistent with an electron hypothesis is
  required.
\item
  Events stemming from interactions of a beam electron with 
  the residual gas or with the wall of the beam pipe
  are not included in the 
  Monte Carlo simulation and are rejected by the requirement
  that  
  the event vertex lies within a cylinder defined by
  $|z_{\rm{ vertex}}| = 10$~cm and having a radius of $3$ cm.
\end{itemize}

After the preselection, $363$ events remain in the sample while 
$339.8 \pm 6.9$ events are predicted by Monte Carlo simulation. 
The contribution of the signal as predicted by grc4f is
$84.2 \pm 1.7$ events. The errors are statistical.

The overall signal efficiency of the preselection predicted by grc4f is
$(27.0 \pm 0.5) \%$. 
Splitting up the events into two different kinematic regions defined by the
invariant mass \mqq\ reveals a
dependence of the efficiency on the event topology.
In the low mass region with
an invariant mass \mqq\ between $5$~GeV and $60$~GeV the efficiency
is $(21.5 \pm 0.6) \%$. Here the main loss in efficiency is due 
to the multiplicity cut.
For \mqq\ $ \ge 60$~GeV the efficiency is
$(33.8 \pm 0.7) \%$. 
Using the PYTHIA generator similar efficiencies are observed.

The efficiencies and especially the differences in the efficiencies in the 
two different mass regions can be understood by 
looking more closely at the topology.
The \Zg\ is predominantly scattered in the forward direction.
Therefore in the low invariant mass region (\gee)
many particles from the hadronic decay stay in the beam pipe, leading to the
loss in efficiency due to the multiplicity cut. On the other hand the high
invariant mass region (\zee) is not affected, as the decay products gain enough
transverse momentum to be detected within the detector.
The  differential distribution of the scattering angle of the electron is 
peaked in the backward direction especially for the high invariant mass 
region, 
strongly reducing the acceptance of the electron.
Consequently requiring one electron to be detected in the central jet chamber
rejects many signal events.
The geometrically accepted region for the outgoing electron
is determined by the minimum number of 
hits required in the jet chamber corresponding to 
\mbox{$|\cos\theta_{\rm e}| \le 0.963$}.

\begin{figure}[h]
  \centerline{\epsfig{figure=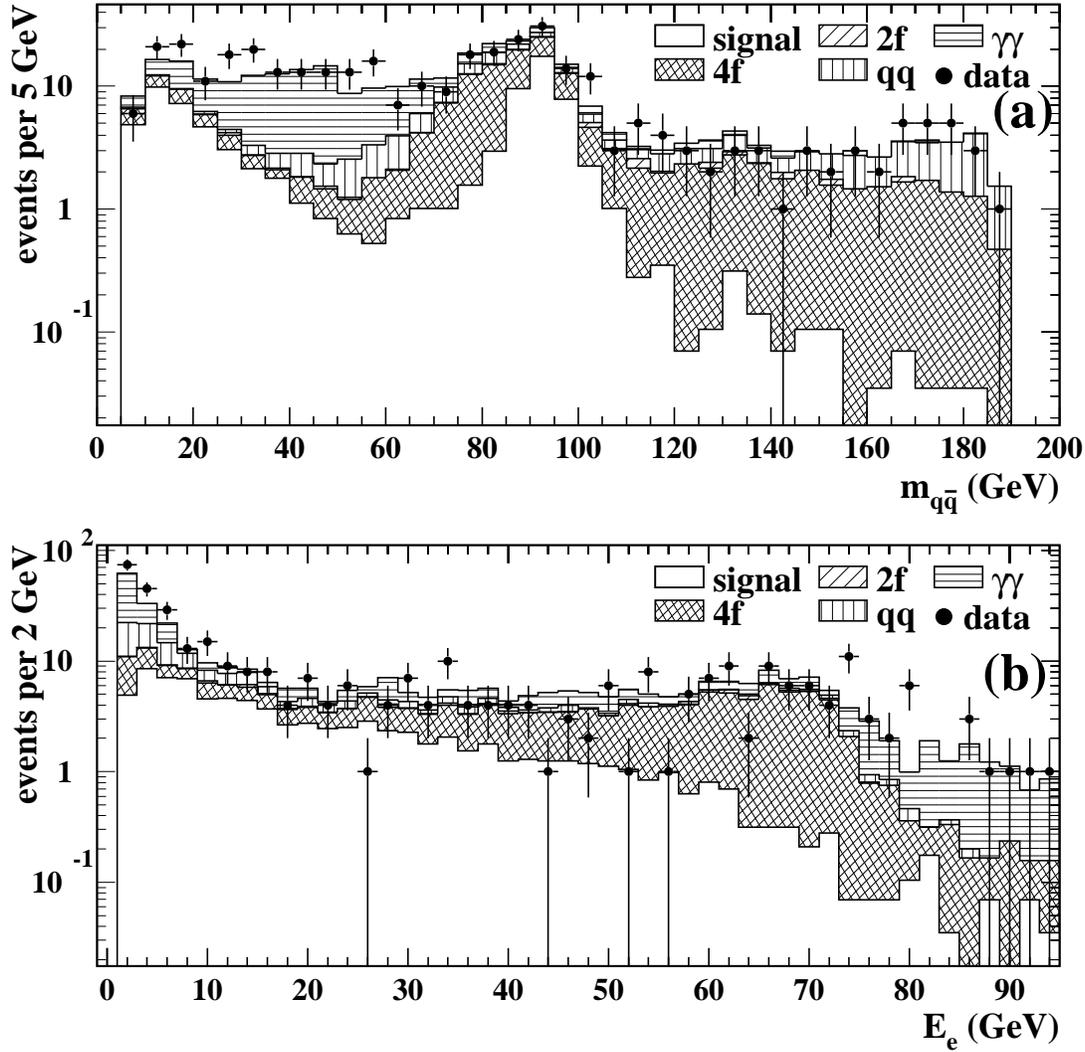,width=1.\textwidth}}
  \caption{\it 
    Distribution of (a) \mqq\ and (b) the electron energy \Elecal\ after the 
    preselection. The histograms show the contributions from the various 
    processes and the points represent the data. 
    The signal contribution is taken from the grc4f Monte Carlo sample.
    Only statistical errors are shown.}
 \label{fig:mqqpre}
\end{figure}

The measured distributions of \mqq\ and
\Elecal\ after the preselection are compared to the Monte Carlo expectations 
in Figure~\ref{fig:mqqpre}.
The distributions are well described by the Monte Carlo simulation.
The resolution of \mqq\ obtained from the kinematic
fit is about $3$~GeV.

The main contribution to the background in the low mass region comes from
electrons from photon conversions in two-photon events. 
In the high mass region, processes with
an electron from semi-leptonic  W$^{\pm}$ pair decays dominate. 
In both regions there is a contribution from falsely identified electrons.
In the signal processes the selected electron candidate is almost always the
scattered beam electron.
Also for tagged two-photon and other four-fermion processes
mostly correctly identified electrons are found.
The tagged two-photon
events often satisfy our kinematic signal definition and are difficult to
separate from the signal process.

The preselection has been improved with respect to the one applied 
in~\cite{zee} by 
including a neural network to identify photon conversions, lowering the minimum
energy requirement for the electron to $1$~GeV and a changed isolation
criterion. The implementation of the kinematic fit improves the resolution in
the quantities describing the hadronic system, leading to a
better description of the 
kinematic variables \mqq, $\sqrt{\hat{s}}$, $\hat{u}$ and $\cos\theta^*$.

\section{Selection of the signal}
\label{selection}
\begin{figure}[h]
  \centerline{\epsfig{figure=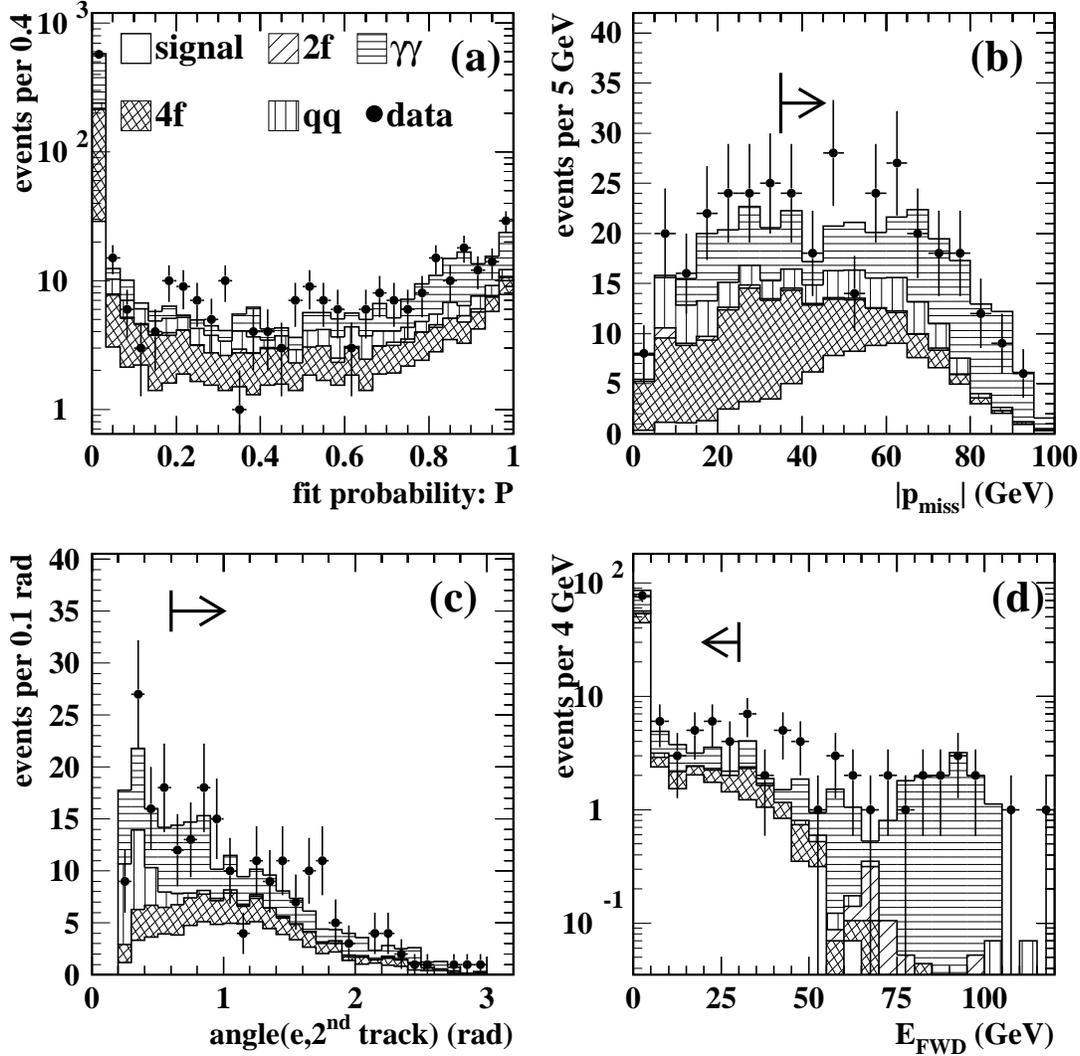,width=16cm}}
 \caption{\it 
   Distributions of variables used in cuts in the preselection and 
   in the selection.
  (a)  $P$, the fit probability,
   (b) $p_{\mathrm{miss}}$, the missing momentum,
   (c) the angle between the electron track and the nearest track,
   and (d) $E_{FWD}$, the energy deposited in the forward calorimeter
   before applying the cut on that variable.
   The arrows indicate the
   selected region.
   Only statistical errors are shown.
   The histograms show the contributions
   from the various  
   processes and the points represent the data. 
   }
 \label{fig:sel}
\end{figure}

After the preselection, the ratio of signal to background is approximately 
$1$ to $3$. In order to  further reduce the background, the following cuts 
are applied. 
The distributions of the variables 
used in  each cut  are shown in Figure~\ref{fig:sel}. 
The numbers of events after each cut are shown in Table~\ref{tab:sel} 
for data, Monte Carlo signal and the various background processes.
\begin{table}[htb]
\tabcolsep0.1cm
\begin{center} 
\begin{tabular}{|l||r|r|r|r|r|r|c|} \hline
&\multicolumn{6}{|c|}{Number of expected events from MC}& OPAL \\ 

\rb{Cut}
&\multicolumn{1}{|c}{\zgee} 
&\multicolumn{1}{c}{4f} 
&\multicolumn{1}{c}{$\gamma\gamma$} 
&\multicolumn{1}{c}{\qq}
&\multicolumn{1}{c}{2f}
&\multicolumn{1}{c|}{Sum}  
&\multicolumn{1}{|c|}{data} \\ \hline \hline
        Presel. & $   84.2\pm    1.7$ & $   100.0\pm    1.8$ & $  106.9\pm    6.3$
 & $  44.9\pm    1.2$ & $   3.8\pm    0.6$ & $ 339.8\pm    6.9$ & $ 363 $\\ \hline
        Cut1 & $   71.2\pm    1.6$ & $   39.9\pm    1.2$ & $   80.9\pm    4.2$
 & $  23.8\pm    0.9$ & $   1.2\pm    0.3$ & $ 217.1\pm    4.7$ & $ 224 $\\ \hline
        Cut2 & $   59.1\pm    1.4$ & $   29.5\pm    1.0$ & $   54.2\pm    3.2$
 & $   3.3\pm    0.3$ & $   1.2\pm    0.3$ & $ 147.4\pm    3.7$ & $ 154 $\\ \hline
        Cut3 & $   57.6\pm    1.4$ & $   14.2\pm    0.7$ & $   53.6\pm    3.2$
 & $   3.2\pm    0.3$ & $   0.7\pm    0.2$ & $ 129.3\pm    3.6$ & $ 140 $\\ \hline
        Cut4 & $   53.8\pm    1.4$ & $   11.4\pm    0.6$ & $   35.0\pm    2.5$
 & $   2.7\pm    0.3$ & $   0.4\pm    0.1$ & $ 103.2\pm    2.9$ & $ 101 $\\ \hline
        Cut5 & $   48.1\pm    1.3$ & $    8.3\pm    0.5$ & $    9.0\pm    1.3$
 & $   2.4\pm    0.3$ & $   0.3\pm    0.1$ & $  68.1\pm    1.9$ & $  70 $\\ \hline
\end{tabular} 
\end{center}
\caption{\it Numbers of expected and observed events for an integrated
  luminosity
  of $174.7 \; \ipb$ ~after each cut. The number of expected
  signal events is obtained using the grc4f generator. The numbers of
  background events are evaluated using the Monte Carlo samples
  described in the text. The errors are statistical only. }
\label{tab:sel}
\end{table}

\begin{figure}[h]
\vspace{-2.5cm}
  \centerline{
    \epsfig{figure=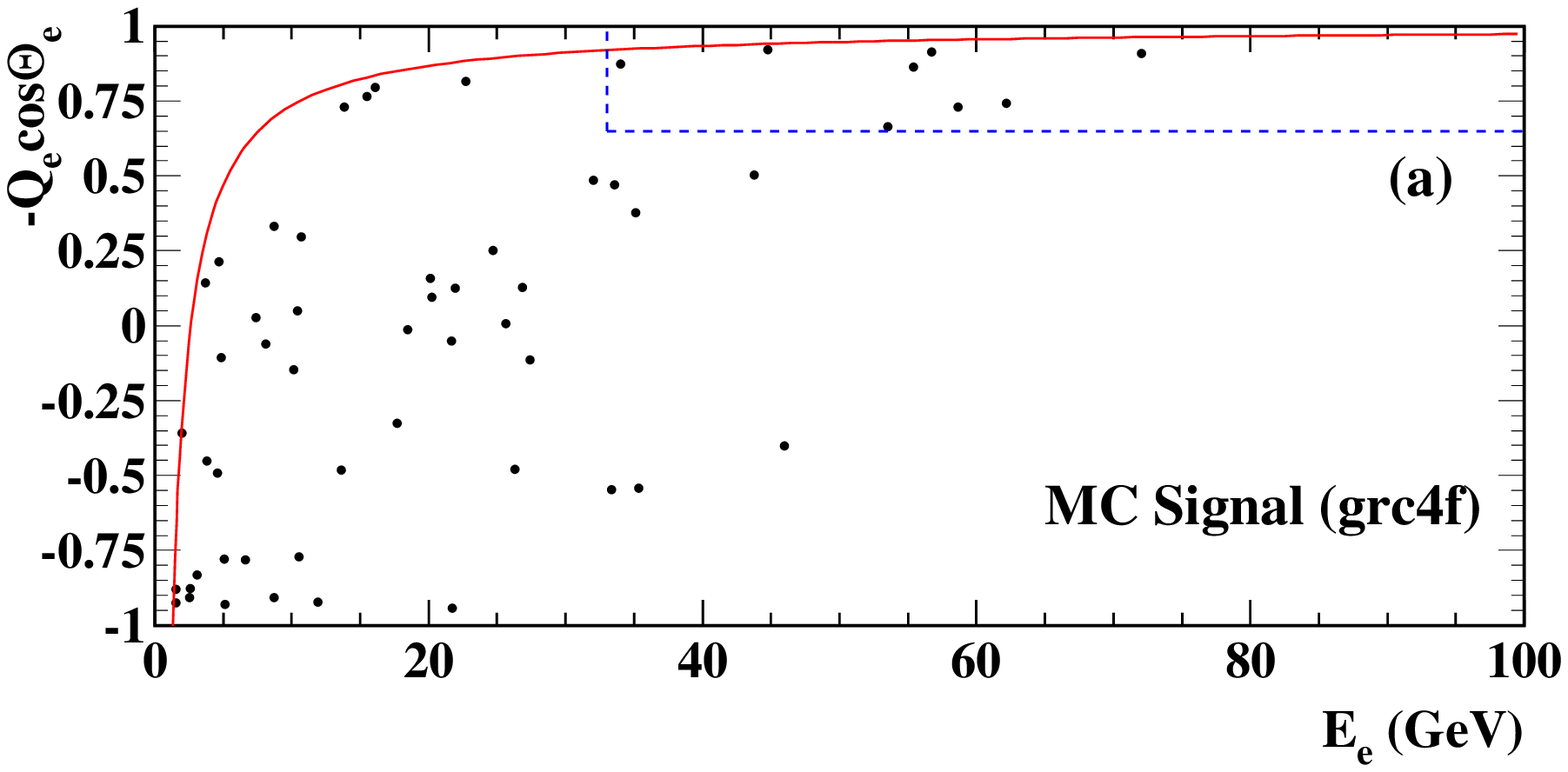,width=15cm}}
    \label{fig:ethetasig} 
\vspace{-8cm}
  \centerline{
    \epsfig{figure=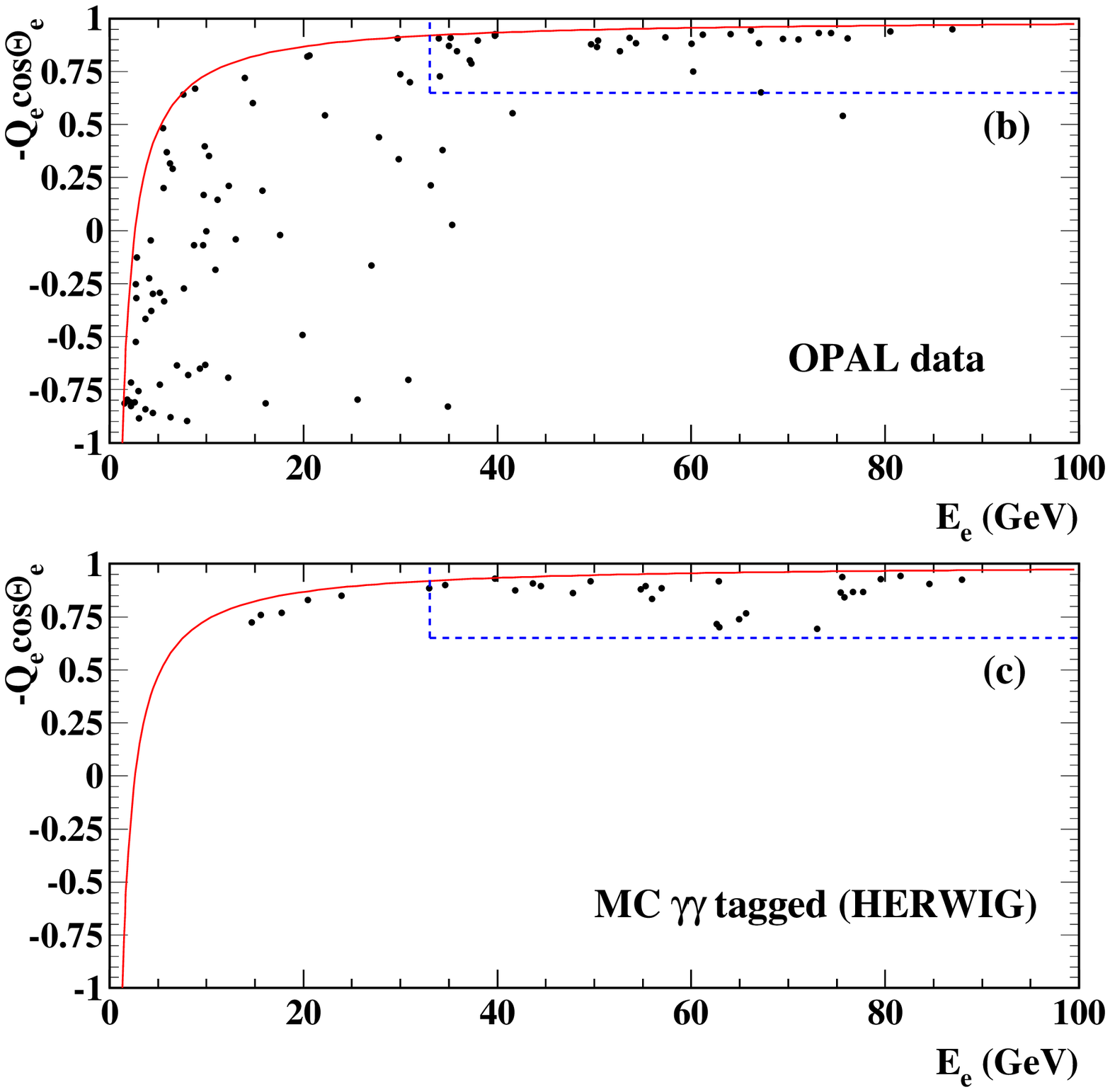,width=15cm}}
    \caption{\it Distribution of the cosine of the scattering angle of the
      electron \elcosth\ versus its energy \Elecal\ before the last cut. 
      In (a) the distribution is shown for the signal, in (b) for the data and
      in (c) for the tagged two-photon Monte Carlo sample.
      Both Monte Carlo samples have been scaled to the same 
      integrated luminosity as the data.
      The solid line indicates the cut in the preselection at 
       $|\hat{t}|\ge 500 \; \gev^2$. The area inside the dashed line 
      corresponds to the region
      discarded by the last selection cut.}
    \label{fig:etheta} 
\end{figure}

\begin{description}
\item[(Cut 1)] The absolute value of the missing momentum must fulfil 
  $\pmiss \ge 35$~GeV. In the signal events the
  missing momentum is due to the electron which emitted the quasi-real photon
  and remains in the beam pipe.
\item[(Cut 2)] To reduce the background from multi-hadronic events 
  the isolation criterion for the signal electron is tightened,
  requiring that the angle 
  between the electron and the second closest track be at least $0.6$~rad.
\item[(Cut 3)] For the signal the missing momentum points in the direction 
  of the electron staying inside the beam
  pipe, for this reason the missing momentum vector of the event must satisfy
  $\qe \cosmiss \ge 0.95$.
\item[(Cut 4)] The maximum energy allowed in the forward detectors 
  $E_{FWD}$ is $30$~GeV. 
  With this cut the remaining events from the two-photon process
  where one electron is tagged by the forward detectors are reduced.
\item[(Cut 5)] In order to remove the remaining background from tagged 
  two-photon reactions those events are  rejected
  where the scattering angle of the electron is in the forward
  direction by requiring $-\qe \elcosth \le 0.65$ or 
  $\Elecal \le 0.35 E$. This cut is illustrated in 
  Figure~\ref{fig:etheta}.

\end{description}

After all cuts, $70$ events are selected while $68.1 \pm 1.9$ events
are expected from the Monte Carlo prediction, of which $48.1 \pm 1.3$ are
signal.  This corresponds to an overall signal efficiency of $(15.2 \pm 0.4)
\%$ according to grc4f.
The main contribution to the background stems from tagged two-photon events
where one of the scattered electrons is detected within the detector.
In the region removed by the last cut the events from two-photon
processes are dominant, as shown in Figure~\ref{fig:etheta}. For the region
outside this cut, the \zgee\ events are dominant, but still a non negligible
contribution from tagged two-photon events remains.

The cross-section \see\
is measured in two different regions of \mqq, 
in the low invariant mass region \gee\ and in the high invariant
mass region \zee. By separating the two mass regions 
at a point where the measured 
cross-section is near its minimum, the expected feed-through,
i.e.\ the number of events with a true value of \mqq\ outside 
the region it is measured in, is very small.
The results of the cross-section measurement for both grc4f and PYTHIA 
are summarised in 
Table~\ref{tab:xsec}. 
The efficiency for the high mass region is about 50\%  
larger than that for the low
mass region and the expected number of signal events is similar. 
In the low mass 
region the background is higher, stemming mainly from tagged two-photon events.

\begin{table}[htb]
\begin{center} 
\begin{tabular}{|l|c|c||c|c|} \hline

& \multicolumn{2}{|c||}{grc4f} 
& \multicolumn{2}{|c|}{PYTHIA} \\
\hline 
& \gee 
& \zee 
& \gee 
& \zee \\
\hline 
\hline

Efficiency  in $\%$
& {$ 13.2 \pm 0.5 $}
& {$ 18.0 \pm 0.6 $} 
& {$ 14.4 \pm 0.7 $}
& {$ 19.4 \pm 0.8 $} \\
\hline
 
Expected signal
& {$ 22.7 \pm 0.9 $}
& {$ 25.4 \pm 0.9 $} 
& {$ 26.0 \pm 1.5 $}
& {$ 30.2 \pm 1.6 $} \\
  
Expected background
& {$ 12.1 \pm 1.1 $}
& {$  7.9 \pm 0.8 $} 
& {$ 12.1 \pm 1.1 $}
& {$  7.9 \pm 0.8 $} \\
  
Feed through in \mqq
& {$ 0.1 \pm 0.1 $}
& {$ 0.2 \pm 0.1 $} 
& {$ 0.1 \pm 0.1 $}
& {$ 0.3 \pm 0.2 $} \\
  
OPAL data
& {$ 40 $}
& {$ 30 $} 
& {$ 40 $}
& {$ 30 $} \\
\hline
 
Measured cross-section in pb
& {$ 1.20 \pm 0.28 $}
& {$ 0.69 \pm 0.18 $} 
& {$ 1.11 \pm 0.26 $}
& {$ 0.64 \pm 0.17 $} \\
\hline
 
Predicted cross-section in pb
& {$ 0.98 \pm 0.01 $}
& {$ 0.80 \pm 0.01 $} 
& {$ 1.03 \pm 0.02 $}
& {$ 0.88 \pm 0.02 $} \\
\hline 
\end{tabular} 
\caption{\it Comparison of OPAL data with the cross-section \see\ 
predicted by the grc4f and the PYTHIA
Monte Carlo generator. The errors are statistical only.}
\label{tab:xsec}
\end{center}
\end{table}

\begin{figure}[h]
 \centerline{\epsfig{figure=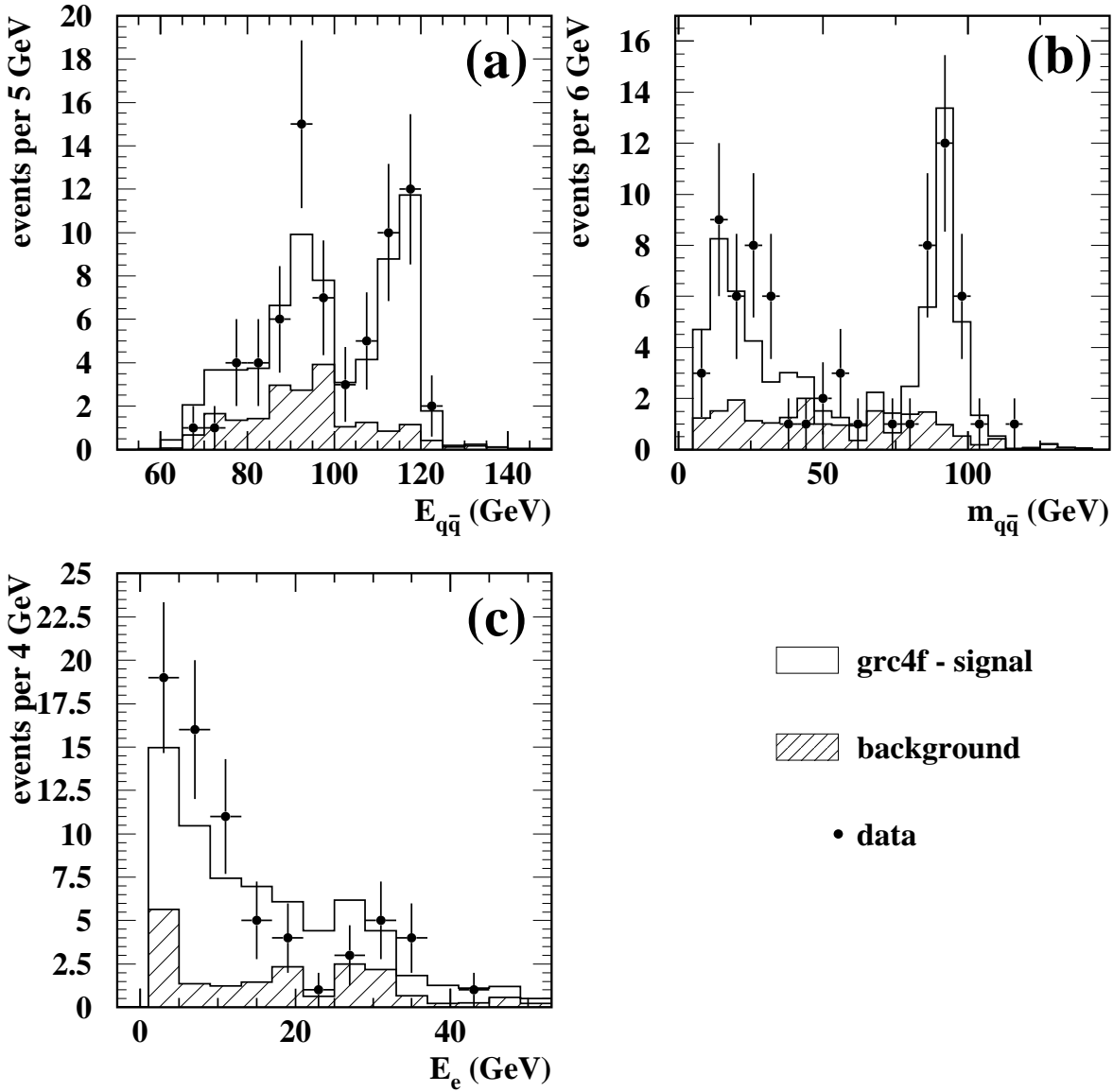,width=16cm}}
 \caption{\it 
   Number of events measured after applying all cuts as a function of
   (a) the energy \Eqq\ and (b) mass \mqq\ of the quark system and of   
   (c) the electron energy \Elecal. 
   The open histogram shows the signal simulated with the grc4f Monte
   Carlo, the hatched histogram shows the backgrounds
   from various Monte Carlo simulations
   and the points  the data. 
   Only statistical errors are shown.
  }
 \label{fig:mqqelecal}
\end{figure}

\begin{figure}[h]
 \centerline{\epsfig{figure=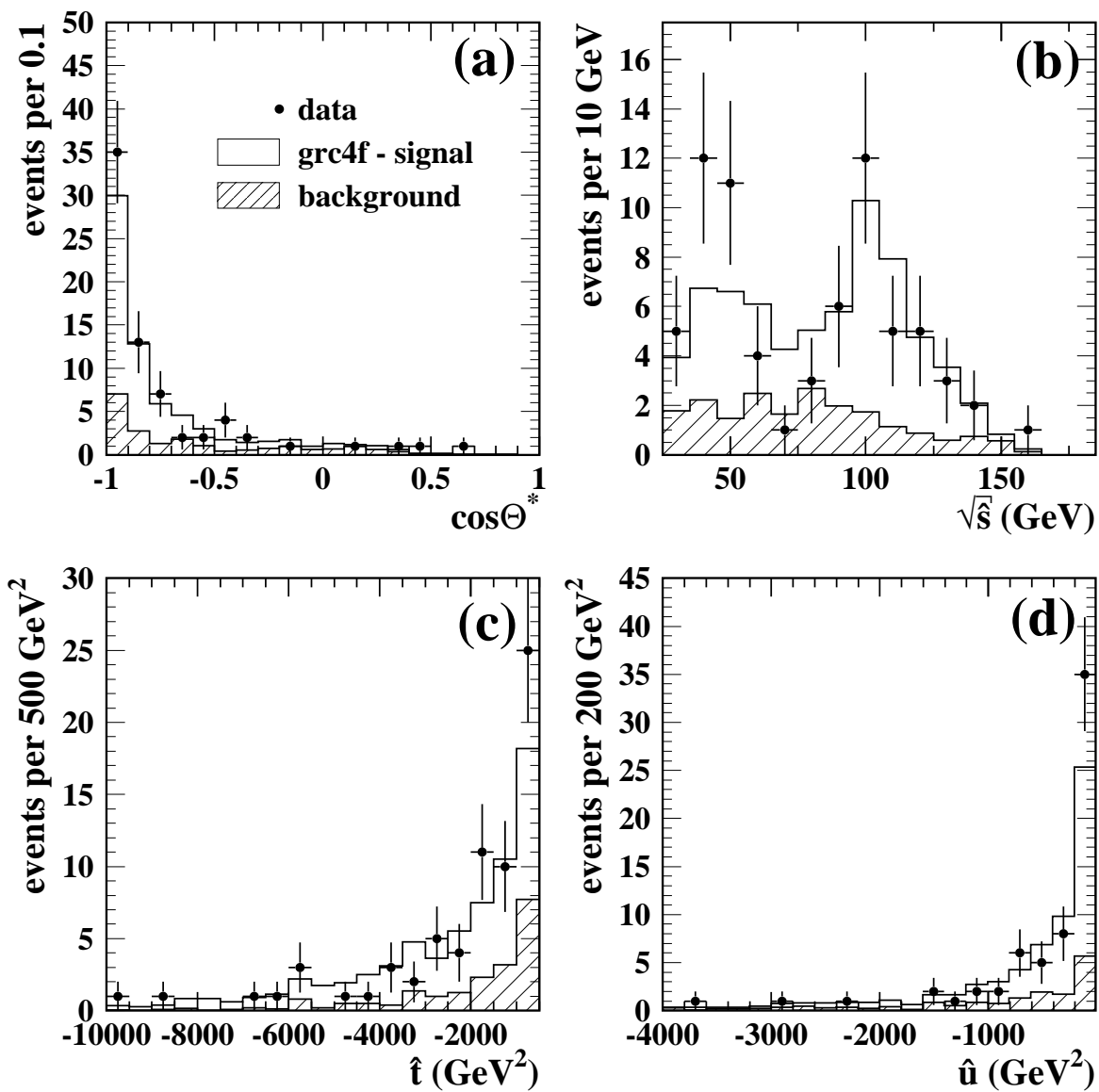,width=16cm}}
 \caption{\it 
  Distribution of (a) the scattering angle in the e$\gamma$ rest-frame
   $\cos\theta^*$, and 
   the kinematic invariants (b) $\sqrt{\hat{s}}$, (c) $\hat{t}$ and 
   (d) $\hat{u}$    after all cuts. 
   The open histogram shows the signal simulated with the grc4f Monte
   Carlo, the hatched histogram shows the backgrounds
   from various Monte Carlo simulations
    and the points  the data. Only statistical errors are
   shown.}
 \label{fig:t}
\end{figure}

\begin{figure}[h]
 \centerline{\epsfig{figure=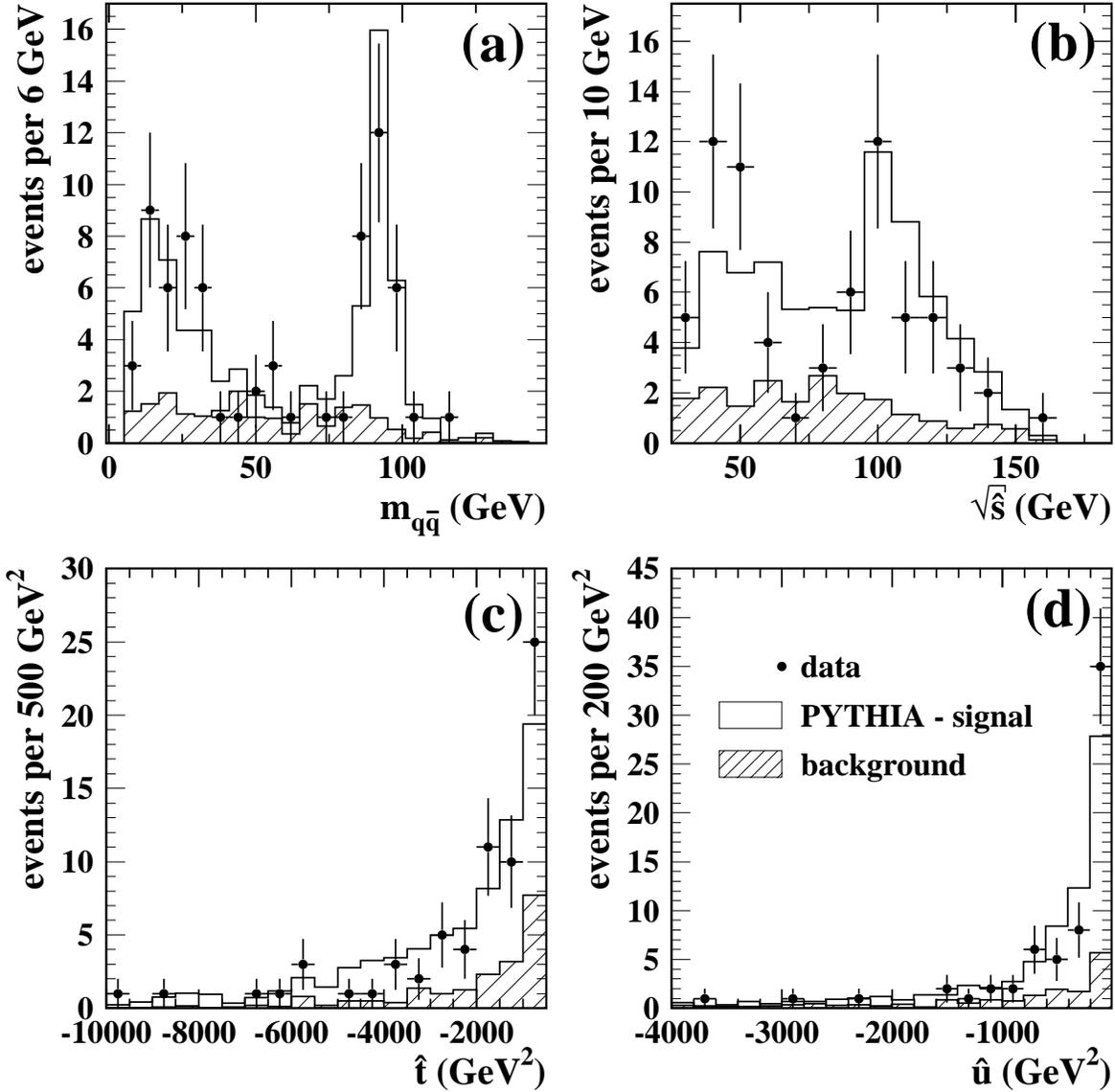,width=16cm}}
 \caption{\it 
  Distribution of (a) the mass \mqq\ and 
   the kinematic invariants (b) $\sqrt{\hat{s}}$, (c) $\hat{t}$ and 
   (d) $\hat{u}$    after all cuts. 
   The open histogram shows the signal simulated with the PYTHIA Monte
   Carlo, the hatched histogram shows the backgrounds
   from various Monte Carlo simulations
    and the points  the data. Only statistical errors are
   shown.}
 \label{fig:pythia}
\end{figure}

In Figures~\ref{fig:mqqelecal} and ~\ref{fig:t}
the measured event distributions for several variables are compared to
the ones predicted from the grc4f Monte Carlo simulations.
For comparison the event distributions from PYTHIA
are given in Figure~\ref{fig:pythia} for some variables.
The limited statistics of the data does not allow to distinguish
between the two simulations.

\section{Determination of differential cross-sections}
\label{sec:diffsec}

To compare the results of this analysis with results from 
other experiments, differential cross-sections are calculated from the
distributions of the event variables. 
For the observed process 
differential cross sections d\see\ and d\seg\
have been determined.

{\boldmath \subsection{Differential cross-sections d\see }}
\label{sec:sigmaee}

For the determination of the differential cross-sections 
the experimental resolution is of importance.
If the experimental
resolution is much smaller than the chosen bin width and the distribution is 
flat, then the correlation matrix will be close to the unit matrix. But if the
distribution is peaked, like for \mqq\, 
then a large fraction of events measured in bins around the peak 
originated from bins other than the one they had been generated in.
Therefore the correlation for each variable between the generated 
and the measured distribution has to be determined. 

The correlation matrix $M$ between the generated 
and the measured distribution has been calculated for each variable shown in 
Figures~\ref{fig:mqqelecal} and \ref{fig:t} using the bin width shown there.
The matrix $M$ is given by:
\begin{equation}
M(i,j)= \frac{G(i,j)}{\sum_jG(i,j)}
\label{eq:matrix}
\end{equation}
and  fulfils the following condition:
\begin{equation}
N_{gen}(i) = \sum_{j} M(i,j) N_{det}(j),
\label{eq:ngen}
\end{equation}
where $N_{gen}(i)$ is the number of events generated in bin $i$,
$N_{det}(j)$ is the number of events measured in bin $j$ and
$G(i,j)$ is the number of events generated in bin $i$ and measured in bin $j$.
The matrix $M$ has been determined using the grc4f MC. For most
variables $M$ is very similar to the unit matrix with some small 
off-diagonal elements. For \mqq\ around the Z-mass non zero elements exist 
also away from the first off-diagonal. 

The matrix $M$ has also been calculated from the PYTHIA signal MC and no 
significant difference with the one determined from grc4f has been observed.

\begin{figure}[h]
 \centerline{\epsfig{figure=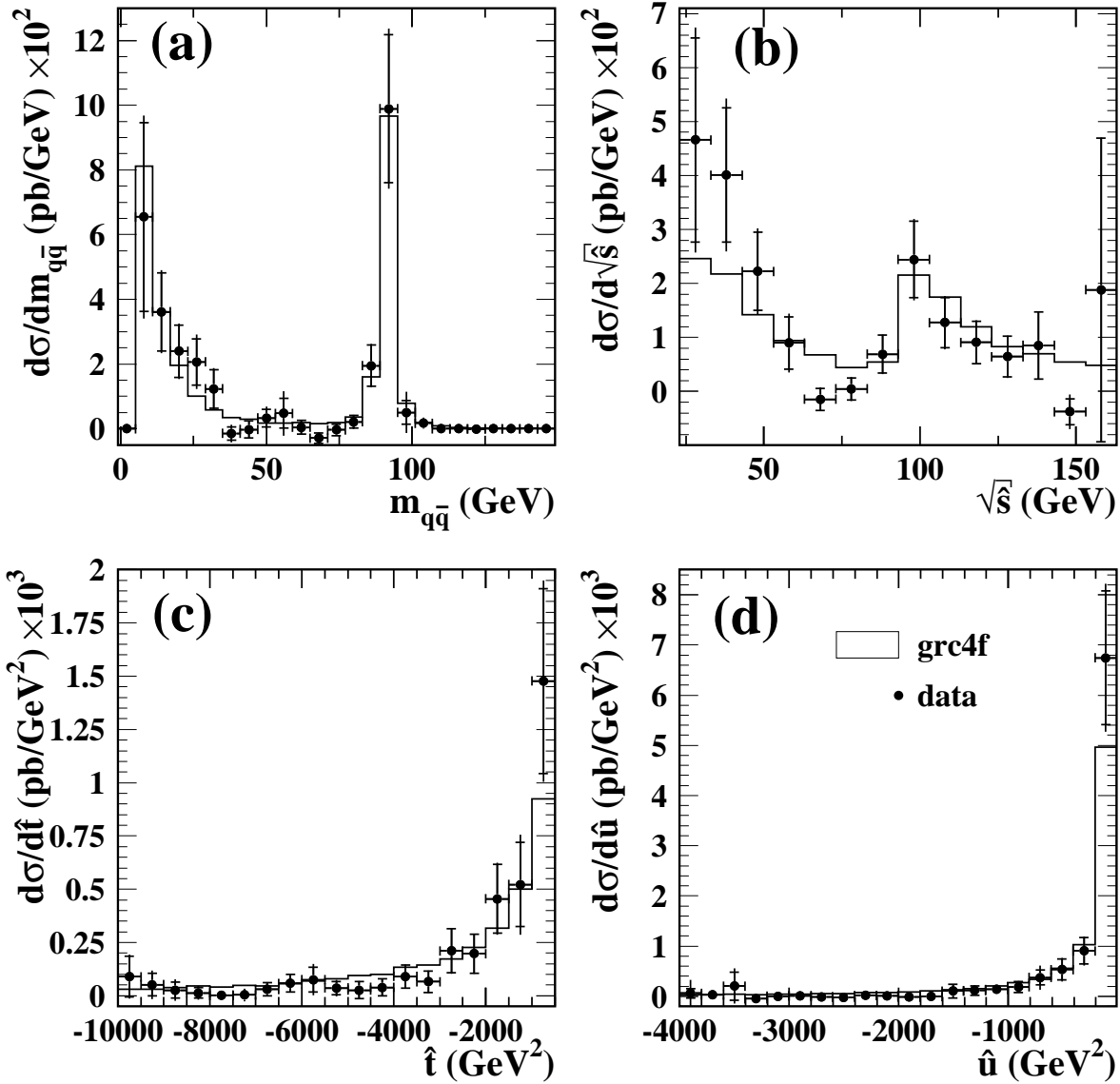,width=16cm}}
 \caption{\it 
   The differential cross-sections 
(a)   $\mathrm{d}\sigma_{ee}/\mathrm{d}m_{\qqs}$,
(b)   $\mathrm{d}\sigma_{ee}/\mathrm{d}\sqrt{\hat{s}}$,
(c)   $\mathrm{d}\sigma_{ee}/\mathrm{d}\hat{t}$ and
(d)   $\mathrm{d}\sigma_{ee}/\mathrm{d}\hat{u}$ are shown.
   The open histograms show the signal simulated with the grc4f Monte
   Carlo generator  and the points the data. The errors show the 
   statistical and systematic errors added in quadrature.
   The contribution of the statistical errors is indicated by the
   horizontal bars.
   }
 \label{fig:diff2}
\end{figure}

The differential cross sections are then given by

\begin{equation}
\frac{{\rm  d} \sigma_{\rm ee}}{{\rm d}\,x} = 
\sum_j M(i,j) \left( N_{det}(j,x) - N_{back}(x) \right) 
\frac{1}{ \Delta x}
\frac{1}{{\cal L}_{\rm ee} \;\,\epsilon(x)} \ ,
\label{eq:sigee}
\end{equation}
where $x$ is the variable used, $ {\cal L}_{\rm ee}$ the integrated 
luminosity and $\epsilon(x)$ the efficiency in a given $x$ bin.
The differential cross-sections are shown in Figure~\ref{fig:diff2}.

{\boldmath \subsection{Differential cross-sections d\seg }}
\label{sec:sigmaeg}

A further aim of this analysis is to calculate the 
differential cross-sections d\seg.
This allows the results from
a given \ee\ centre-of-mass energy to be compared 
with other energies as well as 
with results from other colliders.
To calculate the differential cross-section for a variable $x$, 
Equation~\ref{eq:sepa} has to be inverted:
\begin{equation}
  \frac{{\rm d}\hat{\sigma}_{{\rm e}\gamma}}{{\rm d} \, x} 
  = \frac{1}{\Delta z} \;\; \frac{1}{\Dge} \;\; \frac{{\rm d}\sigma_{\rm ee}}{{\rm d} \, x},
\end{equation}
with \Dge\ being the  photon-flux and $\Delta z = \Delta \hats / s$.
The lower limit of $\sqrt{\hats}$ is given by the signal definition
as 23~GeV and the upper limit is  160~GeV, resulting in 
$\Delta z = 0.702$.
To calculate 
${\rm d}\hat{\sigma}_{{\rm e}\gamma}/{\rm d} \, x $
the mean of the inverse of the photon-flux  $<1/\Dge>$ is calculated in bins
of \hatu\ and \hats.
Taking into account the dependence of $<1/\Dge>$ on \hatu\ is necessary, 
as the efficiency varies with \hatu.
We have chosen three bins in \hatu\ and four in \hats. The bin boundaries 
as well as the mean values of $<1/\Dge>$ are given in Table~\ref{tab:meandge}.
The dependence of $<1/\Dge>$ on \hatu\ is small while it is large
for \hats.
The width of the bins is chosen to be at least three times larger than 
the experimental resolution.

\begin{table}[h!tb]
\begin{center} 
\begin{tabular}{|l l||c|c|c|} \hline

& &\multicolumn{3}{c|}{bins in \hatu\ in GeV$^2$} \\
& &\multicolumn{3}{c|}{ } \\

\multicolumn{2}{|c||}{\raisebox{2.5ex}[-2.5ex]{$\bigg< \frac{1}{\Dge} \bigg>$ }} 
& $-50000$ to $-1000$
& $-1000$  to $-200$
& $-200$ to $-10$ \\
\hline \hline

bins
& $23$ to $50$
& $0.69$
& $0.67$
& $0.59$ \\
\cline{2-5}

in $\sqrt{\hats}$
& $50$ to $80$
& $2.02$
& $2.03$
& $2.24$ \\
\cline{2-5}

in GeV
& $80$ to $110$
& $6.08$
& $6.61$
& $7.19$ \\
\cline{2-5}

& $110$ to $160$
& $13.90$ 
& $14.49$
& $16.06$ \\
\cline{2-5}

\hline \hline
\end{tabular} 
\caption{\it Mean values of  the inverse of the photon flux 
$<1/\Dge>$ in bins of \hatu\ and $\sqrt{\hats}$. 
} 
\label{tab:meandge}
\end{center}
\end{table}

The differential cross-section d$\sigma_{\rm ee}$/d$x$ 
is calculated according to Equation~\ref{eq:sigee} 
using bins of \hatu\ and \hats.

\begin{equation}
  \frac{{\rm  d} \sigma_{\rm ee}}{{\rm d}\,x} = \sum_{\hatu,\hats} 
  \frac{N(\hats,\hatu,x) / \Delta x}{{\cal L}_{\rm ee} \;\,\epsilon(\hatu,x)}
\ ,
\end{equation}
where $N(\hats,\hatu,x)$ is the number of events in bins of
\hats, \hatu\ and $x$ after subtracting the background
and using the  matrix $M$. 
Larger bins compared to the previous section have been used and $M$
is calculated using these bin sizes.
The efficiency $\epsilon(\hatu, x)$ is calculated only in bins of 
\hatu\ and $x$ as it is flat in $\sqrt{\hats}$.
This results in the differential cross-section in the \eg\ system to be 
given by:
\begin{equation}
  \frac{{\rm  d} \hat{\sigma}_{{\rm e}\gamma}}{{\rm d}\,x} = \sum_{\hatu,\hats} 
  \frac{N(\hats,\hatu,x) / \Delta x}{{\cal L}_{\rm ee} \;\,\epsilon(\hatu,x)} \;\,
  \frac{1}{\Delta z} \;\, \bigg< \frac{1}{\Dge} \bigg> \; .
  \label{eq:diffxs}
\end{equation}
The measured differential cross-sections
are shown in Figure~\ref{fig:dall}
and are compared to the generated distributions.

For calculating the total cross-section \seg$(\sqrt{\hats})$ 
as a function of $\sqrt{\hats}$ the same method as above is applied,
with the difference that  $\Delta z$ is calculated  for
each bin of $\sqrt{\hats}$.
\section{Systematic error studies}
\label{syserrors}

For the calculation of the total and differential cross-sections 
the efficiencies, unfolding matrix and backgrounds are taken from
Monte Carlo simulations. It is therefore important to study systematic
effects resulting from these simulations.

\subsection{Systematic error studies for the total cross-section}

The systematic errors on the efficiencies come mainly from imperfect modelling
of the detector response.
This can lead to discrepancies between the data and the Monte Carlo simulation 
in the distributions of the cut variables.
These systematic errors can be estimated, for example, by comparing Monte
Carlo simulation and data.

This has been done using the events selected by the preselection.
For these events each of the selection cuts has been applied
separately and the relative difference in the number of events selected in data
and in Monte Carlo has been assigned as a systematic error
after quadratically subtracting the statistical error. In cases where 
this results 
in a value being lower than the statistical error, conservatively the
statistical error is used.
An error common for the whole mass range has been calculated
and no distinction between the low and high mass range has been made.
The values of the errors are listed in Table~\ref{tab:sys}.
These systematic uncertainties are used for both the \gee-like and \zee-like kinematic 
regions since the efficiencies of each of these selection cuts are similar in
the two regions. 

As a cross check the systematic errors have also been estimated
by comparing Monte Carlo simulation and data for the
process \WWqqen\ which
has the same observable final state as the \zgee\ process.
\WWqqen\ events are selected 
according to the procedure described in \cite{WWQQLN}
and then each selection cut is applied
separately to this sample.
For those cuts where the distribution of the cut variables is similar
for \WWqqen, \zee\ and \gee\ events
(the absolute value of the missing momentum, the
electron isolation and the electron angle and energy)
no difference within the statistical error
in the systematic error compared to the method described above
has been observed.

The relatively smaller overall efficiency for \gee-like 
events arises mainly from the multiplicity cut in the preselection.
To asses the systematic uncertainty of this cut 
the number of tracks and clusters required has been changed by $\pm 1$.

\begin{table}[htb]
\begin{center} 
\begin{tabular}{|l||c|c|} \hline
& \gee & \zee               \\  \hline \hline
multiplicity                            & 0.024 & ---     \\ 
$\pmiss \ge 35$~GeV                     & 0.042 & 0.042      \\ 
electron isolation                      & 0.038 & 0.038      \\ 
$-\qe\elcosth$ cut                      & 0.028 & 0.028      \\ 
$\qe\cosmiss \ge 0.95$                  & 0.033 & 0.033      \\ 
$\Efwd \le 30$~GeV                      & 0.056 & 0.056      \\ 
\hline \hline
detector simulation                     & 0.094 & 0.091      \\ 
efficiency                              & 0.049 & 0.050      \\ 
background                              & 0.038 & 0.033      \\ \hline \hline
Total                                   & 0.113 & 0.109      \\ \hline 
\end{tabular} 
\caption{\it 
Relative systematic uncertainties of the cross-section measurements.
The entry ``detector simulation'' is the quadratic sum of the signal
efficiency uncertainties for the single cuts listed in the rows above
it.}
\label{tab:sys} 
\end{center} 
\end{table} 

The uncertainty in the efficiency due to the choice of a particular 
Monte Carlo event generator
is estimated by comparing PYTHIA and grc4f. 
For this comparison 
only events with the signal electron within the acceptance of the
detector, defined by a cut  
on generator level on the angle of the electron
\mbox{$|\cos\theta_{\rm e}^{\rm{gen}}| \le 0.963$}
are used.
The relative difference in efficiency of the two different 
Monte Carlo generators after subtracting the statistical errors quadratically
is taken as a systematic error.

Uncertainties affecting the residual background have been evaluated
separately for
each of the three main background classes remaining after all cuts.
Background-enriched samples are obtained by inverting or omitting
one or two cuts, while the other cuts remain unchanged. The full difference
between the number of events remaining in the data and the number of expected
events from Monte Carlo is taken as a systematic uncertainty.
For the background from four-fermion final states the cut on the 
fit probability is omitted and the cut on the electron
isolation is inverted. 
After applying these cuts, 
a relative difference
of  $17 \%$ between the data and Monte Carlo is observed.
The background from tagged two-photon events is enriched by inverting 
the cut on the electron's angle and energy. 
The relative systematic error is $10 \%$.
By omitting the cut on the angle of the missing momentum and inverting the
electron isolation cut the multi-hadronic background is enriched, leading to a
relative systematic uncertainty of $12 \%$. 

The numbers of background events after all cuts in the \gee\ region
are 
$4.47 \pm 0.39 \pm 0.75$ from four fermion events, 
$1.07 \pm 0.19 \pm 0.11$ from multi-hadronic processes and 
$5.80 \pm 1.01 \pm 0.73$ events from tagged two-photon reactions.
In the Zee region the corresponding contributions are
$ 3.81 \pm 0.34 \pm 0.64$,
$ 1.38 \pm 0.22 \pm 0.14$ and 
$ 2.46\pm  0.66\pm 0.31 $ events.
Additional background sources contribute with less than
$1$ event for each mass region.
This leads to a relative error on the cross-section of $3.8\%$ in the low mass
region and $3.3\%$ in the high mass region, as quoted in Table~\ref{tab:sys}.

In further studies contributions to the background not modelled in the Monte
Carlo were investigated. Using random beam-crossing events, the
interactions between the beam particles and the 
gas inside the vacuum pipe were found to be negligible.

For the multi-hadronic background an additional systematic cross-check is applied
by comparing the prediction of two different Monte Carlo generators. A good 
agreement between the PYTHIA and YFS3FF Monte Carlo generators has been found
within the statistical errors.
Consequently no additional systematic error has been assigned.

\subsection{Systematic error studies for the differential cross-sections}

The systematic errors for the differential cross-sections stem mainly
from the imperfect detector simulation in the Monte Carlo
and from the uncertainty in the  unfolding matrix $M$.

The error assigned for the 
detector simulation is taken to be the same as for the total cross-section.
A value of 9.4\% is assigned.
The unfolding matrix $M$ is calculated using both
the grc4f and the PYTHIA Monte Carlo sample
and the relative difference between the two is taken as a systematic error.
For the background the same errors as determined for the total 
cross-section are used. The errors for the three different background
classes are taken into account in each bin of the event distributions.
The efficiency $\epsilon$ 
is calculated in bins of the variable $x$ for the
differential cross-sections d\see\ and in bins 
of \hatu\ and the variable $x$ 
for the differential cross-sections d\seg.
It  has therefore  much larger
statistical errors than in the calculation of the total cross section.
Within the statistical errors no difference between the efficiencies
determined from the grc4f and the PYTHIA Monte Carlo samples
has been observed.
Consequently no systematic error has been assigned.

Due to the small statistics of the data sample the systematic
error for the differential cross-sections is much smaller than the statistical
one.

\section{Results and discussion}
\label{conclusions}

The  cross-section for the process \eeZg\ has been measured in a
restricted phase space, defined by Lorentz invariant variables,
in two different regions of the mass of the hadronic system, corresponding 
to either a $\gamma^{*}$ or a Z$^{0}$ in the final state. 
With the cut at \mqq\ = $60$~GeV
the $\gamma^*$ and the Z$^0$ are well separated.
With an integrated luminosity of about $175$~pb$^{-1}$, a total of $70$
candidate events have been observed, while $20.0 \pm 1.4$ background events 
and $48.1 \pm 1.3$ signal events are predicted,  
giving a total of $68.1 \pm 1.9$ events.
The cross-sections 
times branching ratio for the decay of \Zg\ into hadrons, \see,
are found to be
$\sigma = (1.20 \pm 0.28 \pm 0.14)$~pb for \gee\
and 
$\sigma = (0.69 \pm 0.18 \pm 0.08)$~pb for \zee\ final states 
within the kinematical definition 
listed in Table~\ref{tab:sig}.
A large part of this cross-section is not detectable within the 
detector as the scattering angle of the electron is in the very backward 
direction. 
For the calculation of the cross-sections the efficiencies predicted by the
grc4f generator are used. The cross-sections measured using efficiencies 
predicted by PYTHIA lie well within the errors.

The distributions of \Eqq , \mqq and \Elecal\ are shown in 
Figures~\ref{fig:mqqelecal}(a) to (c), respectively. 
The distribution of \Eqq\ 
shows two peaks, one at the beam energy and one at
about $115$~GeV.
From equation~\ref{eq:utlab} one obtains for the largest part of the
cross-section at $\hat{u} = 0$ and $ Q_e\cos\theta_{\qqs} = -1$
\begin{equation}
E_{\qqs} = E_{\rm beam}  + \frac{m_{\rm q \bar{q}}^2}{4 E_{\rm  beam}}.
\end{equation}
Consequently, the peak at the beam energy corresponds to \gee\ 
and the peak at $115$~GeV to the \zee\ process.
The tail at lower energies is due to the parts of the
cross-section where $\hat{u} \ne 0$ and $Q_e\cos\theta_{\qqs} \ne -1$. 
Two peaks are visible in the invariant mass distribution, 
stemming from the contributions of
the two gauge bosons, the \gstar\ and the Z.
The expected background is flat over the whole mass range up to
100~GeV.
For both distributions the shapes of Monte Carlo and data are in good
agreement.
The two Monte Carlo generators give similar distributions and 
more statistics is needed to distinguish between the two.

In Figure~\ref{fig:t} the distributions of 
the scattering angle $\theta^*$ of the 
\Zg\ in the e$\gamma$ rest-system with respect to the incoming 
photon direction,
as well as the distribution of
the kinematic invariants $\sqrt{\hat{s}}$, $\hat{t}$ and $\hat{u}$ are shown
and are compared with the predictions of grc4f.
The scattering angle $\theta^*$ peaks strongly in the backward direction
and the agreement between data and Monte Carlo is good.
The observed structure of the distribution of  $\sqrt{\hat{s}}$,
can be understood in terms of the \gee\ final states in the low  
$\sqrt{\hat{s}}$ region and the \zee\ in the high  $\sqrt{\hat{s}} $ region.
In Figure~\ref{fig:pythia} the measured distributions are compared to
the distributions from PYTHIA.
The predictions of both grc4f and PYTHIA are in agreement with the data. 
The distribution of $\hat{t}$ is peaked towards $0$ and
shows a long tail towards large values.
The distribution of $\hat{u}$ shows the typical behaviour  of
a $u$-channel process, a peak at zero.

From these event distributions the 
differential cross-sections d\see\ are derived and 
are shown in Figure~\ref{fig:diff2}. Only Lorentz invariant quantities
have been derived.
The Monte Carlo describes the data well, except for small values of 
$\sqrt{\hat{s}}$, where the Monte Carlo underestimates the data. 
In the distribution of $\mathrm{d}\sigma_{ee}/\mathrm{d}m_{\qqs}$
for small values of \mqq\ the steep falloff of the 
cross-section with increasing invariant mass \mqq\ 
as well the peak at the Z-mass are very well visible.
The distribution of $\mathrm{d}\sigma_{ee}/\mathrm{d}\sqrt{\hat{s}}$
shows a decrease with increasing 
$\sqrt{\hat{s}}$ until the threshold for Z Boson production
is reached. 

\begin{figure}[h]
 \centerline{\epsfig{figure=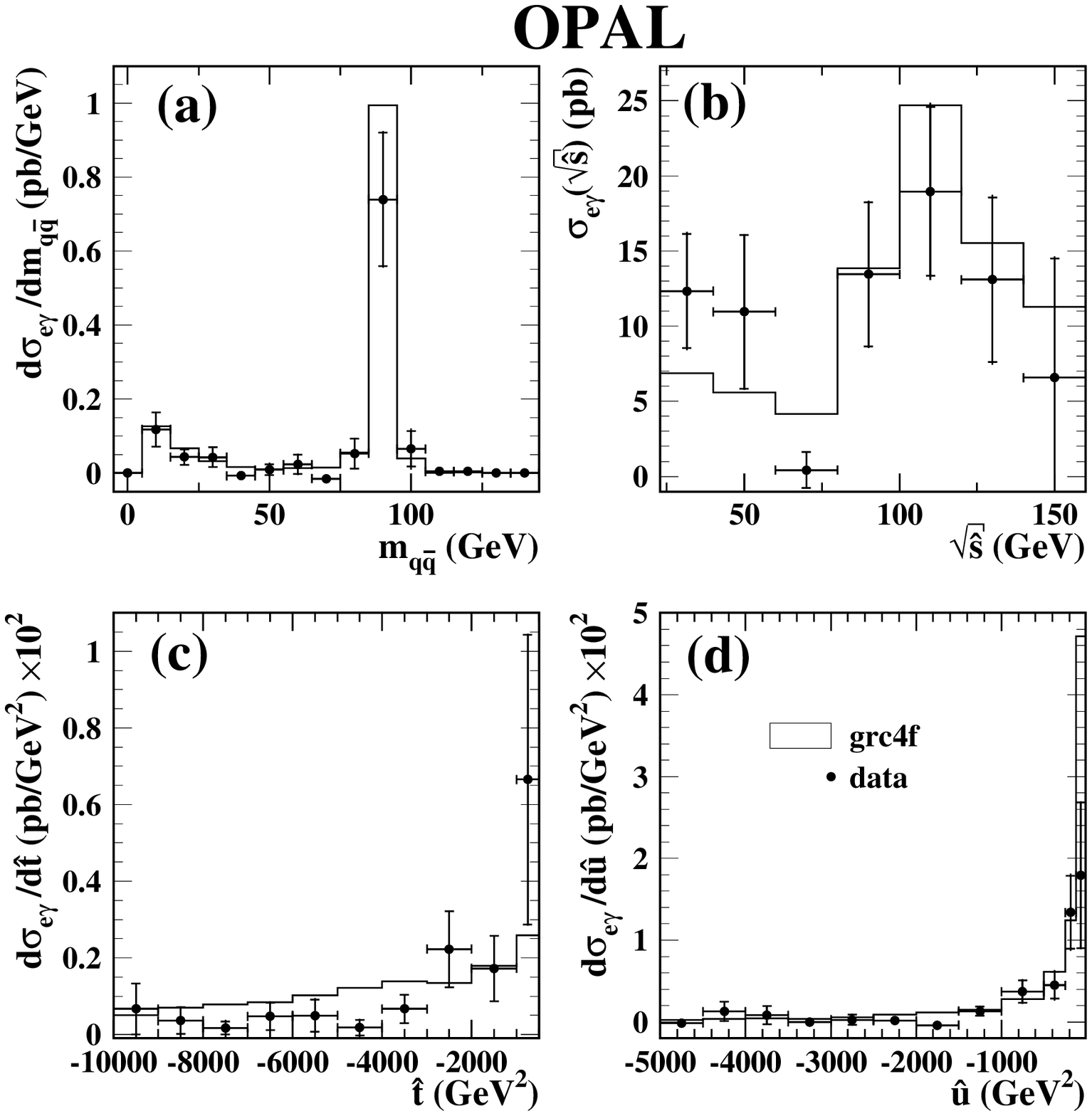,width=16cm}}
 \caption{\it 
   The differential  cross-sections 
(a)   d\seg/$\mathrm{d}m_{\qqs}$,
(c)   d\seg$/\mathrm{d}\hat{t}$ and
(d)   d\seg$/\mathrm{d}\hat{u}$ 
are shown.
In addition (b) the cross section \seg$(\sqrt{\hat{s}})$
is shown
   The open histograms show the signal simulated with the grc4f Monte
   Carlo generator  and the points the data. The errors show the 
   statistical and systematic errors added in quadrature.
   The contribution of the statistical errors is indicated by the
   horizontal bars.
   }
 \label{fig:dall}
\end{figure}

From the differential cross-sections d\see\ 
the differential cross-sections for the sub-process 
\mbox{$\eg \to \mathrm{e}$ \Zg\ }
are deduced, based on a factorisation ansatz using the modified
Equivalent Photon approximation from Equation~\ref{eq:dge}.
They are shown in  Figure~\ref{fig:dall}. 

The distribution of d$\seg / \rm{d}m_{\rm{qq}}$  shows a strong
peak around the Z Boson mass.
The differential cross-section  d\seg/d\hatu\
shows a very sharp peak at 0, as expected for this $u$-channel process.
For d\seg/d\hatt\ the Monte Carlo does not describe the data so well.
The increase for \hatt\ towards 0 is well reproduced, but for large negative 
values of \hatt\ the Monte Carlo lies constantly above the data.

In Figure~\ref{fig:dall}(b) the total cross-section \seg($\sqrt{\hats}$) 
is shown.
Note the dip in the cross-section at $\sqrt{\hats}$ around 70~GeV.
The total cross-section \seg($\sqrt{\hats}$)
shows  a decrease with increasing $\sqrt{\hat{s}}$ until the threshold 
for Z Boson production is reached. 
This is the first measurement of \seg($\sqrt{\hats}$) 
around the Z Boson threshold.
\seg($\sqrt{\hats}$) is independent of the $\ee$
centre-of-mass energy and can therefore be compared with measurements
at other $\ee$ centre of mass energies as well with measurements
at other colliders, e.g. HERA.

Within the statistical error, the Monte Carlo predictions 
are in good agreement with the data.
But there is a tendency that the data 
are higher than the Monte Carlo in the low $\sqrt{\hat{s}}$ region,
while they are too low in the high $\sqrt{\hat{s}}$ region.

\section{Conclusions}

The process \eeZg\ and its subprocess \egezg\ have been studied.
For the process \eeZg\ the cross-section times branching ratio
for the decay of the \Zg\ into hadrons at $\sqrt{s} = 189$~GeV
has been measured
within a restricted phase space to be
$\sigma = (1.20 \pm 0.28 \pm 0.14)$~pb for \mqq $< 60$~GeV and
$\sigma = (0.69 \pm 0.18 \pm 0.08)$~pb for \mqq $\ge 60$~GeV.
The Monte Carlo generators grc4f and PYTHIA both predict 
cross-sections within one standard deviation of the measured values. 
The cross-section \seg($\sqrt{\hats}$)
for the subprocess \egezg\ has been determined
in a range of $\sqrt{\hats}$ from 23 to 160~GeV.

Differential cross-sections d\see\ and d\seg\ have been determined and
compared to the ones from the Monte Carlo generators grc4f and PYTHIA.
The generators describe all distributions well.

\appendix

\noindent
{\Large \bf Acknowledgements}

\noindent
The authors would like to thank T. Sj\"ostrand for his help 
by making the changes to PYTHIA necessary to describe 
the process investigated in this paper.

We particularly wish to thank the SL Division for the efficient operation
of the LEP accelerator at all energies
 and for their continuing close cooperation with
our experimental group.  We thank our colleagues from CEA, DAPNIA/SPP,
CE-Saclay for their efforts over the years on the time-of-flight and
trigger
systems which we continue to use.  In addition to the support staff at our
own
institutions we are pleased to acknowledge the  \\
Department of Energy, USA, \\
National Science Foundation, USA, \\
Particle Physics and Astronomy Research Council, UK, \\
Natural Sciences and Engineering Research Council, Canada, \\
Israel Science Foundation, administered by the Israel
Academy of Science and Humanities, \\
Minerva Gesellschaft, \\
Benoziyo Center for High Energy Physics,\\
Japanese Ministry of Education, Science and Culture (the
Monbusho) and a grant under the Monbusho International
Science Research Program,\\
Japanese Society for the Promotion of Science (JSPS),\\
German Israeli Bi-national Science Foundation (GIF), \\
Bundesministerium f\"ur Bildung und Forschung, Germany, \\
National Research Council of Canada, \\
Research Corporation, USA,\\
Hungarian Foundation for Scientific Research, OTKA T-029328, 
T023793 and OTKA F-023259.\\



\clearpage
\vfill
\vspace{-4cm}
{\Large \bf (a) 
\hspace{8cm}
(b) }

\vfill
%
\vspace{3cm}

\vfill\vfill\vfill


\clearpage

\clearpage


\clearpage

\clearpage


\clearpage


\clearpage


\clearpage


\clearpage


\clearpage


\clearpage

\end{document}